\documentclass[prd,showpacs,amsmath,twocolumn,showkeys,floatfix]{revtex4}
\usepackage{epsfig,dcolumn}
\usepackage{graphicx}
\usepackage[usenames]{color}
\usepackage{graphicx}
\usepackage{bm}
\usepackage{amsfonts}
\usepackage{amsmath}
\usepackage{amssymb}
\usepackage{epsfig}
\usepackage{graphicx}
\usepackage[usenames]{color}

\def\lsim{\mathrel{\rlap{\lower4pt\hbox{\hskip1pt$\sim$}}
    \raise1pt\hbox{$<$}}}
\def\gsim{\mathrel{\rlap{\lower4pt\hbox{\hskip1pt$\sim$}}
    \raise1pt\hbox{$>$}}}

\newcommand{\ar}{\arrowvert}
\newcommand{\bep}{\bm{\epsilon}}

\newcommand{\be}{\begin{equation}}
\newcommand{\ee}{\end{equation}}
\newcommand{\ba}{\begin{eqnarray}}
\newcommand{\ea}{\end{eqnarray}}

\begin{document}

\title{Local Two-Photon Couplings and  the $J=0$ Fixed Pole \\
in Real and Virtual Compton Scattering}
\author{Stanley J. Brodsky}\email{sjbth@slac.stanford.edu}
\affiliation{Theory Group, SLAC National Accelerator Laboratory,
2575 Sand Hill Road, 94025 Menlo Park, California, USA.}
\author{Felipe J. Llanes-Estrada} \email{fllanes@fis.ucm.es}
\affiliation{Depto. F\'{\i}sica Te\'orica I, Fac. Cc. F\'{\i}sicas,
Universidad Complutense de Madrid, 28040 Madrid, Spain.}
\author{Adam P. Szczepaniak}\email{aszczepa@indiana.edu}
\affiliation{Department of Physics and Nuclear Theory Center \\
Indiana University, Bloomington, IN 47405, USA. }

\date{\today}

\begin{abstract}

The local coupling of two photons to the fundamental quark currents of a hadron
gives an energy-independent contribution to the Compton amplitude proportional to the charge squared of the struck quark, a  contribution
which has no analog in hadron scattering
reactions. We show that this local
contribution has a real phase and is universal,  giving  the same contribution  for real or virtual Compton scattering for any photon virtuality and skewness at  fixed momentum transfer squared $t$. The $t$-dependence of this  $J=0$ fixed Regge pole  is
parameterized by a yet unmeasured even charge-conjugation form
factor of the target nucleon.  The $t=0$ limit gives an important
constraint on the dependence of the nucleon mass on the quark mass
through the Weisberger relation.  We discuss how
this $1/x$  form factor  can be extracted from high energy deeply virtual Compton scattering and examine predictions given by models
of the $H$ generalized parton distribution.
\end{abstract}
\maketitle

\section{Introduction and Overview}
\label{intro} Exclusive hadron scattering processes at high energies
are well described by the exchange of Pomeron and  Regge exchanges.
Regge theory, combined with  the vector meson dominance model,
provides a useful description of real and virtual high energy
photoproduction, single-photon processes which at the QCD level
describe photon dissociation into quark-antiquark pairs which
subsequently rescatter off the target constituents. Vector meson
dominance and conventional Regge exchange, however,  cannot account
for contributions to real or virtual Compton scattering where two
photons interact locally on the same quark of the target.

The local coupling of  two photons to the fundamental quark current of a hadron
 leads to  a contribution to the Compton amplitude of the form
\begin{equation}
T^{J=0}_{\gamma^*(q) p \to \gamma^*(q^\prime) p^\prime}=  - 2 e^2
F^{C=+}_{1/x}(t,Q^2)\bep\cdot \bep'. \label{deff}
\end{equation}
The even charge-conjugation, "$1/x$"  form factor $ F^{C=+}_{1/x}(t)$ is real for spacelike $t$. Unlike normal Regge exchange, contributions to the Compton amplitude which behave as $\beta_R(t) s^\alpha_R(t)$, the $J=0$ fixed pole contribution is {\it energy independent}  at any fixed
$t= (q^\prime -q)^2 = (p -p^\prime)^2$.
Remarkably  $T^{J=0}$  is also independent of the incident and final photon virtualities $q^2$
and $q^{\prime 2}$ as well as the skewness  $\xi =-{ (q^2+q^{\prime 2})/ 4p\cdot q }$  for any given fixed $t$.   It thus appears in real photon scattering as well as virtual Compton scattering.
Because it has a real phase, the local contribution to virtual Compton scattering has maximal interference with the Bethe-Heitler bremsstrahlung contributions to  $\ell p \to \ell^\prime \gamma  p$~\cite{Brodsky:1971zh}.  These amplitudes can also be measured in timelike two-photon processes such as
$\gamma^* \gamma^* \to H \bar H$ and $\gamma^*  \to H \bar H \gamma.$
Because the $J=0$ contribution  arises from the local interactions of the two photons, there is no analog in any hadron scattering amplitude,  and thus it cannot be obtained from models based on vector meson dominance.  Unlike normal Regge trajectories, the local two-photon interaction only couples to scalar mesons in the $t$ channel, not a  sum over states with progressively higher orbital angular momentum.  The isospin of the contributing scalar mesons can be
$I=0, 1,$ and $2$. (However, no isospin-2 meson, an exotic by
necessity, is currently well established).

In the case of the proton target, there are two $C=+$ amplitudes with the local
$J=0$ structure: helicity-conserving and helicity flip, analogous to the Dirac and Pauli form factors. The  $ F^{C=+}_{1/x}(t)$  form factor, for each quark flavor is obtained by summing over all quarks in the hadron weighted by $\sum e^2_q$.
The integrand also contains an extra factor of $1/ x$ relative to the Dirac and Pauli
electromagnetic form factor where $x$ is the usual light-front fraction
$x= k^+/ p^+ = (k^0+k^z)/(p^0+p^z)$ of the quark in the hadron light-front wavefunction. Hence the name
``$1/x$'' form factor. It can also be related
 to the $1/x$ moment of the $H$ and $E$ Generalized Parton  Distributions (GPDs),
 which parameterize deeply virtual Compton scattering~\cite{Diehl:2003ny,GolecBiernat:1998ja}.
The helicity-conserving form factor at $t=0$ is the $1/ x$ moment which appears
in the Weisberger formula~\cite{Weisberger:1972hk} for $\partial M^2/\partial m^2_q$.
The two-photon form factors obtained from extracting the $J=0$ contribution
to Compton scattering thus give new complimentary information  on the structure of the target hadron.

\begin{figure}
\includegraphics[height=1.7in]{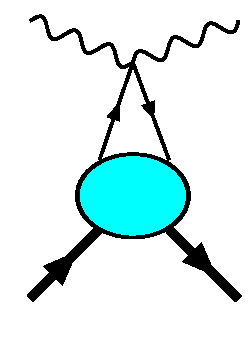}
\caption{\label{fig:seagull}
The local coupling of two photons to a quark yields a fixed pole,
a real and constant contribution to the
Compton amplitude. In scalar electrodynamics
this fixed pole is easy to recognize --it originates from the seagull coupling.
For spin $1/2$ quarks, one needs to pick up  an instantaneous
(on the light-front) Z-diagram component of the handbag diagram,
as shown in Eq.(\ref{fixedpoleinprops}).
}
\end{figure}

The origin of the $J=0$ contribution is trivial in supersymmetric
QCD where the electromagnetic current couples to charged  scalar
squarks.  In this case the theory contains a local four-point
interaction $e^2_s s^\dagger A^\mu A_\mu s$ where $s$ is the squark
field. See Fig.~\ref{fig:seagull}.  This ``seagull'' interaction
couples the two photons locally to the hadron and gives a Born $J=0$
contribution to the Compton amplitude proportional to the charged
squared of the squark.  Since the four-point interaction is local,
all radiative corrections from the strong interactions of the
squarks are incorporated into the $C=+$ hadron form factors. The
fact that a fixed pole can exist in the Compton amplitude $\gamma
p\to \gamma p$ (real or virtual) due to the seagull interaction was
first pointed out by Creutz~\cite{Creutz:1973zf} .

In the case of spin-1/2 quarks, it is convenient to use the
light-front Hamiltonian formulation of QCD~\cite{Brodsky:1997de}.
The seagull interaction with scalar quarks is then replaced by the
light-front instantaneous four-point interaction of the two photons
with the quark current $e^2_q  \bar \psi \gamma \cdot
 A(\gamma^+/ i \partial^+ ) \gamma \cdot A \psi .$  This interaction
arises when one eliminates the constrained
  $\psi^ - =  \Lambda^{-} \psi$ quark field in light-cone gauge $A^+=0$.
The same local two-photon interaction also emerges from the usual
handbag Feynman diagram for Compton scattering.  The numerator of
the quark propagator $\gamma \cdot k_F + m$ appearing between the
two photons in the handbag contributions to the Compton amplitude
contains a specific term $ \gamma^+ \delta k^-/2$  which cancels the
$k^2_F-m^2$ Feynman denominator, leaving a local term inversely
proportional to $k^+$, equivalent to the light-front Hamiltonian
contribution. Here $\delta k^- \equiv k^-_F- (k^2_\perp + m^2)/k^+ =
(k^2_F-m^2)/k^+$  (see Eq.(\ref{spincancel}) below). Thus in the
spin-$1/2$ case, the two-photon interaction is local in impact space
and light-front time $\tau = x^+ = x^0+x^3, $  but it is nonlocal in
the light-front coordinate $\sigma = x^- = x^0-x^3.$
The $J=0$ contribution is intrinsic to the Feynman propagator; it is
also essential for the gauge invariance of the Compton amplitude.

The $J=0$ fixed pole contribution is well known in atomic physics
since it gives the dominant contribution to high energy elastic
Compton scattering on an atom. In this case  the seagull coupling
$e^2 \vec A \cdot \vec  A^\dagger \phi\phi^\dagger$ to the
nonrelativistic electron field of QED is responsible for  the
point-like Thomson scattering on bound atomic electrons.

In principle, the $J=0$ contribution to Compton scattering measures the
local coupling of photons to the fundamental  carriers of the electromagnetic current
at any photon resolution since it is independent of photon energy and virtuality.
However, at finite energies the local contribution is screened by the contributions
to the Compton amplitude from the remaining non-local interactions, For example,
when the photon energy  $\nu/M_p = (s-M_p^2)/2 M_p $ vanishes,
there is no suppression from energy denominators and the nonlocal interactions give the contribution
\begin{equation}
T^{NL}_{\gamma p \to \gamma p}(\nu \to 0) = -2
    \left[   \langle \sum_q e_q \rangle^2  -
    \langle \sum_q  \frac{e^2_q}{x_q} \rangle \right]\bep\cdot \bep^\prime  \label{NL}
\end{equation}
where the sum is over the quarks in the target and   $x_q$ is the
longitudinal momentum fraction carried by the quark of charge $e_q$ (in units of the proton electric charge).
Thus at zero photon energy $s-M^2_p \to 0$, the local contribution,
\begin{equation}
T^{L}_{\gamma p \to \gamma p}( \nu \to 0)    =  - 2
       \langle \sum_q  \frac{e^2_q}{x_q} \rangle  \bep\cdot \bep^\prime  \label{L}
\end{equation}
is exactly canceled, and only the square of the total charge $e_H =
\sum e_q$  appears in the Compton amplitude, $T=T^{L}+T^{NL}$:
\begin{equation}
 T_{\gamma p \to \gamma p}(\nu \to 0)  =
 - 2\langle \sum_q e_q  \rangle^2  \bep\cdot \bep^\prime,
 \end{equation}
consistent with the low energy theorem.  This cancelation of the local and
nonlocal terms is demonstrated explicitly in Ref.~\cite{Brodsky:1968ea}.
Conversely, at higher energies, the nonlocal terms become suppressed or are strongly
modified by the presence of energy denominators; this  in principle, allows
the local terms and their fundamental structure to emerge.

In the case of QCD,  the contribution of the local coupling of the
photons to quarks can be screened by the contributions from  Pomeron
exchange and other $C=+$ Reggeons which have intercepts $\alpha_R(0)
> 0$.  These contributions, which have a complex phase dictated by
analyticity and $t \leftrightarrow u $ crossing,  will typically
dominate the Compton amplitude at high energies. However, the
energy- and $q^2$-independent local contribution proportional to
$\sum e^2_q$  is always present at any scale.  For example, one can
remove the Pomeron by measuring isovector channels such as the
proton/neutron difference.    At large $t$, the non-singlet $C=+$
Reggeons recede to negative values, $\alpha_R(t) < 0$ , exposing the
$J=0$ contribution.  In fact, the $s$-independent contribution
$R_V(t)$ determined at Jefferson Laboratory~\cite{Danagulian:2005vm}
from the elastic Compton scattering amplitude at large $t$, can be
identified with the $J=0$ contribution.  If this identification is
correct, the $R_V$ contribution to the Compton amplitude will be
independent of photon virtuality at fixed  large $t$ in $\gamma^* p
\to \gamma p$ scattering.

In this paper we advocate the study of Compton scattering in the
kinematical regime $s>>-t > -t_0$ in order to extract the $J=0$
fixed pole contribution as a fundamental test of QCD, and to measure
the $1/x$ form factor of the nucleon which parameterizes its $t$
dependence. In addition,  if one studies deeply virtual Compton
scattering, in the kinematical regime $s>>Q^2>>-t >-t_0$, one can
express the form factor in terms of the Generalized Parton
Distribution function $H(x,\xi ~\sim Q^2/s \to 0, t)$, and learn
about its $1/x$ moment. The extrapolation to the forward limit $t
\to 0^-$ provides an important connection to the Weisberger
relation, discussed in Section~\ref{Weisbergersection}. We also note
that the local two-photon coupling  plays an implicit role in all
inclusive processes involving two photons (or other vector fields)
scattering or annihilating on a quark line such as $\gamma^* q  \to
\gamma q$ and $\gamma \gamma \to q \bar q$.

The remaining of the paper is structured as follows.
Section \ref{analyticcont} gives a thorough analysis of the fixed pole
and its parton interpretation. There we comment on point-like scattering
(Sec. \ref{leadingDVCS}), doubly virtual Compton scattering
(Sec. \ref{2VCS}), singly virtual Compton scattering (Sec.
\ref{dvcs}) and real Compton scattering (Sec. \ref{sec:rcs}).

The forward limit of the fixed pole is presented in Section
\ref{sec:forward}, where we examine the range of values suggested by
parton distribution functions in Sec. \ref{subsec:pdfs}. The
approach we follow is  motivated by the procedure of Ref.
~\cite{Brodsky:1973hm} where a scalar-quark model, constrained to
satisfy scaling relations and current conservation was  applied to
describe both real and imaginary parts of the
 $\gamma^* p \to \gamma^* p$ two-photon amplitude.
In  Sec. \ref{Weisbergersection} we review the Weisberger relation
and the connection between the $1/x$ moment of pdf's and the
quark-mass dependence of the nucleon mass. At finite momentum
transfer, the momenta of the two-photons differ and the $1/x$ form
factor can be written as a  moment of the Generalized Parton
Distribution. In Section \ref{sec:simplemodels} we examine a few
simple models of GPDs and conclude that the fixed pole is a general
feature, revealed already by the valence part (three-quark
component) of the nucleon's wave function, independently of whether
traditional Regge theory is or not incorporated in a model. The
common representation in terms of double distributions is briefly
recalled in Sec. \ref{Radyushkin}. Then we provide an estimate of
the $F_{1/x}(t)$ form factor with valence quark model light-front
wavefunctions in Sec. \ref{subsec:lightcone}. We examine the
representation of the GPD in terms of the parton-proton scattering
amplitude at large $t$ in Section \ref{sec:parton-proton}, and give
a corresponding estimate for $F_{1/x}(t)$.

Section \ref{sec:expext} is dedicated to a preliminary examination
of existing data sets.  Early studies gave partial evidence for the
pole in the forward  \cite{Damashek:1969xj} and in the off-forward
\cite{Shupe:1979vg} Compton amplitudes. We show that it is unlikely
that current exclusive data has convincingly revealed the fixed-pole
behavior, opening possibilities for the $12\mbox{ GeV}$ Jefferson
Laboratory facility  or a proposed electron-ion collider.
Conclusions and outlook are then presented in section
\ref{sec:last}. We give a brief primer on the novel analytic
properties of the  fixed pole in Appendix \ref{intropole}. The
implications of isospin symmetry are reviewed in Appendix
\ref{sec:isospin}.

\section{ Parton model interpretation of the $J=0$ pole in two-photon processes  }
 \label{analyticcont}

\subsection{Parton model and point-like scattering}
\label{leadingDVCS}

The appearance of a $J=0$ pole in hadronic Compton  processes originates from the
local coupling of   two photons with the quark constituents of the target hadron.
We define the generalized Bjorken scaling variable
\begin{equation}
\xi \equiv - \frac{q^2 + q'^2}{4\nu}
\end{equation}
where $\nu = (s-u)/4$. In the case of doubly virtual Compton
scattering $q^2 = q'^2 = -Q^2$ and  $\xi$ becomes the Bjorken
variable -$x_B$, known from deep inelastic scattering (DIS), $x_B
\equiv Q^2/(2p\cdot q)$ with $p$  being the  momentum of the target.
In the case of singly virtual Compton scattering $q^2 = -Q^2$ and
$q'^2=0$, $\xi$  becomes the symmetric scaling variable used in Ref..~\cite{Ji},
and in this case it is related to the Bjorken variable -$x_B$: $\xi
= x_B/(2-x_B)$.
The physical interpretation of the $J=0$ pole is easiest to address
in terms of the light-front coordinates $a^\mu = (a^+,a^-,a_\perp)$
with $a^\pm = (a^0 \pm a^3)$ in the frame
 where the $+$ component of the incoming photon
  momentum  vanishes, $q^+=0$.  In this frame the light-front energy of the
quark which is exchanged between the photons is given by either $k^- + q^- $
and $k'^- - q^-$ for the $s$ and $u$ channel amplitude, respectively,
 both being of the order of  $Q^2/\xi$. Thus the $\xi \to 0$ limit
corresponds to the situation where the exchanged quark does not
propagate over the light-front time.

Unlike the spin-0 case, where the seagull contribution is explicitly local
 in all four space-time directions, the high  energy limit
   of a spin-1/2 exchange  extends over the $t-z$,
direction conjugated to the  longitudinal moment $k^+$ and $k'^+$.
 In the Bjorken limit  virtual Compton amplitude, ($q^2 = -Q^2$, $q'^2 =0$) is proportional to
\begin{eqnarray} \label{spincancel}
\frac{\not k + \not q + m}{(k+q)^2-m^2+i\epsilon} \to
\frac{\gamma^+}{2p^+}\left( \mathbf{\frac{1}{x}} +
\frac{\xi}{x}\frac{1}{x-\xi+i\epsilon} \right)  \nonumber \\
= \frac{\gamma^+}{2p^+} \frac{1}{x-\xi+i\epsilon}
 \nonumber \\
- \frac{  \not k - \not q' + m}{(k-q')^2-m^2+i\epsilon} \to
\frac{\gamma^+}{2p^+}\left( \mathbf{\frac{1}{x}} -
\frac{\xi}{x}\frac{1}{x+\xi-i\epsilon} \right) \nonumber \\
= \frac{\gamma^+}{2p^+} \frac{1}{x+\xi-i\epsilon}  \nonumber \\ \label{fixedpoleinprops}
\end{eqnarray}
for direct and crossed-handbag respectively, which are shown in Fig.~\ref{handbag}.
Here $x$ is the  fraction of the incoming proton longitudinal  momentum
  carried by the   struck quark, $x = k^+/p^+$.
 In the right-hand side of each term we show the $1/x$ piece
which comes from canceling the $q^-$ between numerator and
denominator. As can be seen, upon taking the high-energy limit $\nu
\to 0$, ($\xi \to 0$) the remaining terms (from the on-shell
numerator of the hard-quark) cancel out as they are proportional to
$\xi$.  One can immediately infer that in the high-energy limit both
real and virtual Compton scattering on a quark contains the
equivalent of the scalar seagull diagram for Dirac spin-$1/2$
fermions. This contribution reduces,  at fixed $Q^2$, to $ 1/x$, the
fixed pole for elementary fermions and subsequently to the universal
$1/x$ form factor of the target.

\subsection{$J=0$ pole in Forward Spatial  Doubly-Virtual Compton Scattering}
\label{2VCS}
\begin{figure}
\includegraphics[height=2.5in]{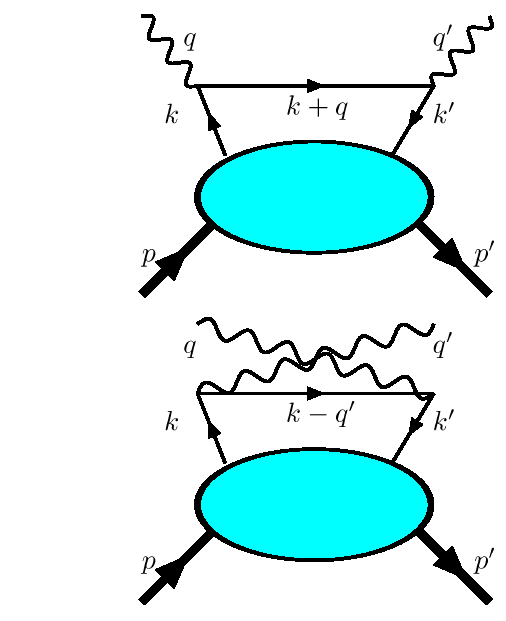}
\caption{\label{handbag}
The handbag and crossed-handbag diagrams convoluting the hard scattering
amplitude  together with a soft Generalized Parton
Distribution.
}
\end{figure}
The manner in which the  $J=0$ pole manifests itself in the real and
virtual Compton amplitude depends on the result of convolution with
the parton distribution of the target hadron. The contribution from
the $J=0$ pole to the doubly-virtual Compton amplitude, $\gamma^*(q)
p \to \gamma^*(q') p'$,  where both the initial and final state
photons are spacelike, has been extensively studied in the past.
Here we summarize the main results following
 Refs.~\cite{Zee:1972nq,Creutz:1973zf}

In the forward case, {\it i.e.} $q=q'$, $p=p'$ doubly  virtual Compton amplitude for a transverse photon,
$\gamma^*(q) p \to \gamma^*(q) p$, $T_1(Q^2,\nu)$,
\begin{equation}
T_{\gamma^*(q) p \to \gamma^*(q) p} = \bep \cdot \bep'  T_1(Q^2,\nu)
\end{equation}
with  $Q^2 = -q^2$,
$\nu =  (s-u)/4 = p\cdot q$. $T$ is a an analytical function of
$\nu$ except
for cuts running along the real axis and starting at $\pm \nu_{th}$.
These assumptions and Cauchy's theorem  lead to a dispersion
representation,
\begin{equation}
T_1(Q^2,\nu) =   C_{\infty}  +
\frac{1}{\pi} \int_{\nu^2_{th}}^{\infty} \frac{d\nu'^2}{\nu'^2 -
\nu^2 -i\epsilon} \mbox{Im}T_1(Q^2,\nu')   \label{disp}
\end{equation}
where $C_\infty$ is the possible contribution from a part of the integration
contour  at infinity in the complex $\nu$ plane. Here  we also used the crossing,
$s-u$ symmetry relation  $T_1(Q^2,\nu-i\epsilon ) = T_1(Q^2, -\nu+i\epsilon)$.

The spectral function $F_1(Q^2,\nu)  = (1/\pi) \mbox{Im}T_1(Q^2,\nu)$
is nonzero for $2\nu \ge 2\nu_{th} = (M_N + m_\pi)^2 + Q^2 - M_N^2 $.
For real Compton scattering, with $Q^2=0$  it is convenient to subtract
the dispersion relation at $\nu=0$ which eliminates
 the contribution from $C_\infty$  and replaces it by the known value
of the amplitude $T_1(0,0) = -2$ --the Thomson term.
In the case of the forward doubly virtual Compton amplitude we can determine $C_\infty$
in the Bjorken limit.  In terms of the Bjorken variable  $x_B = Q^2/2\nu$,
$T_1(x_B) = \lim_{Q^2\to \infty} T_1(Q^2,\nu)$,
 Eq.(\ref{disp}) becomes ($x \equiv Q^2/2\nu'$),

\begin{equation} \label{T1forwardCS}
T_1(x_B) = \Delta +  T_{1,handbag}(x_B)
\end{equation}
where we defined
\begin{equation}
T_{1,handbag}(x_B) \equiv  \int_0^1 dx \frac{2x}{x_B^2 - x^2 - i\epsilon} f(x)
\end{equation}
and
\begin{equation}
\Delta = T_1(x_B) - T_{1,handbag}(x_B) = C_\infty + 2\int_0^1 \frac{dx}{x} f(x)
\end{equation}
Here and in the following $f(x)   = (1/\pi) \mbox{Im }T_1(x)$ is the structure
 function.  This structure function represents the sum over quark and anti-quark
 distributions weighted with the parton's charge
\begin{equation}
f(x) = e_q^2 \left[ f_q(x) + f_{\bar q}(x)  \right]
\end{equation}
 The contribution to $T_1$ from the parton model is
represented by the handbag diagram of Fig.~\ref{handbag} and
 is precisely given by $T_{1,handbag}$ as discussed in
Sec. \ref{leadingDVCS}.
In Fig.~\ref{handbag} the blob represents strong parton-nucleon
interactions and the upper part represents  the hard scattering of
virtual photons off  a free   quark (we ignore non-leading twist and
perturbative QCD corrections).

In general, from phenomenological considerations, as well as from
QCD evolution, it is expected that at small-$x$ the structure
function is given by
\begin{equation}
 f(x) \to f_R(x) =  \sum_\alpha f^\alpha_R(x) = \sum_{\alpha}\frac{\gamma_\alpha}{x^\alpha} \label{sf}
 \end{equation}
  with $\alpha$  both positive and negative being the intercept of the Regge trajectory,
  $\alpha = \alpha(t=0)$;  Pomeron exchange with $\alpha \sim 1$ and $t$-channel meson
  Regge trajectories with $\alpha \sim 0.5$ are clearly visible in the
data. Contributions from daughter trajectories and/or valence quarks
typically have $\alpha<0$. A physical  interpretation of the Regge contributions to hadron structure functions is discussed in Ref.~\cite{Brodsky:1990gn}.

Furthermore, in the Bjorken limit,  deeply inelastic scattering is best interpreted
as if the parton model gives the entire contribution to $T_1$,
{\it i.e.}
\begin{equation}
\Delta=0
\end{equation}
It thus follows that the  Regge contribution to $T_1$ is given by,
   \begin{eqnarray}
  &&T_{1,R}(x_B)  =    \int_0^1 dx \frac{2x}{x_B^2 - x^2 - i\epsilon} f_R(x)  =  \nonumber \\
   & = & \pi  \sum_{\alpha} \frac{1 + e^{i\pi\alpha}}{\sin\pi\alpha}
   f^\alpha_R(x_B)
    +     T^{J=0}_{1,R}   +  T^{J<0}_{1,R}(x_B) \nonumber \\ \label{t1regge}
 \end{eqnarray}
 where
 \begin{eqnarray}
 T^{J=0}_{1,R} &= &  2 \sum_{\alpha}  \frac{\gamma_\alpha}{\alpha}  \nonumber \\
 T^{J<0}_{1,R}(x_B)  & = & 2 \sum_{\alpha } \sum_{J=-2,-4,\cdots}
 \frac{\gamma_\alpha}{\alpha - J} \frac{1}{x^J_B} \label{J}
\ .
\end{eqnarray}
Note that there is no constant contribution to the structure function since the phase of a $\alpha=0$ contribution to the forward Compton amplitude  is real.

We have grouped terms according to their  importance at high energies
(small-$x_B$).  The $J=0$ contribution corresponds to the $T_{1,R}^{J=0}$ term and  $T^{J<0}_R$
represents subleading terms at high energies.

Even though  there is a $J=0$ pole at the level of elementary
interaction between photons and quarks, originating from the $1/x$
term in the photon-quark scattering amplitude, it could be the case
that the convolution of the elementary amplitude with the structure
functions  removes the $J=0$ pole  in the full amplitude. This could
happens if
  \begin{equation}
 \sum_{\alpha}  \frac{\gamma_\alpha}{\alpha} = 0
 \end{equation}
Thus point-like interactions are  necessary but not sufficient for
the existence of a $J=0$ pole.  However, since the Regge
contributions are $t$-dependent. $\alpha= \alpha(t)$, this
accidental cancelation would only occur at one value of $t$.

For a general structure function $f(x)$, the small-$x$
behavior is carried by the Regge-type
$f^{\alpha > 0}_R(x)=\sum_{\alpha > 0} \gamma_\alpha/x^\alpha$
that we have already analyzed, such that
\begin{equation}
\lim_{x\to 0} \left[ f(x) - f^{\alpha >  0}_R(x) \right] =0  \ .
 \label{interpol}
\end{equation}
After isolating the Regge and non-Regge parts the $T_1$ amplitude can be written as,
\begin{eqnarray}  \label{2vcsfinal}
T_1(x_B)  & = & \pi  \sum_{\alpha >  0} \frac{1 + e^{i\pi\alpha}}{\sin\pi\alpha} f^{\alpha>0}_R(x_B)
+ T^{J=0}_1 +  T_1^{J<0}(x_B)
 \nonumber \\
\end{eqnarray}
where
\begin{eqnarray}  \label{2vcsfinal2}
T_1^{J=0} & = & - 2 \int_0^1 \frac{dx}{x} f_v(x)  + 2\sum_{\alpha>0} \frac{\gamma_\alpha}{\alpha}
\nonumber \\
T_1^{J<0}(x_B) & = &   x_B^2 \int_0^1 \frac{dx}{x} \frac{2}{x^2_B - x^2 - i\epsilon} f_v(x)  \nonumber \\
 &+ &  2\sum_{\alpha>0} \sum_{J=-2,-4,\cdots}
\frac{\gamma_\alpha}{\alpha - J}\frac{1}{x_B^J}
\end{eqnarray}
and we have defined the valence structure function as
\begin{equation}
f_v(x) \equiv f(x)  - f^{\alpha > 0}_R(x) \label{valence}
\end{equation}
so that  by construction $f_v(0) = 0$. It thus follows that at high
energies $T^{J>0}_1(x_B\to 0)$ vanishes. In the forward direction it
is the first term on the {\it r.h.s.} which dominates and grows with
energy as $1/x_B^\alpha$.

\subsection{ Non-forward, Spatial Doubly Virtual Compton Scattering}

The $T_1$ amplitude at finite momentum transfer $t = (p'-p)^2$,
$\gamma^*(q) p \to \gamma^*(q') p'$ can be obtained from analyticity
in $t$, which implies analyticity in $\alpha$ since $\alpha=
\alpha(t)$, $f(x) \to f(x,t)$ (which will be interpreted shortly as
a Generalized Parton Distribution)  with
\begin{equation}
f^\alpha_R(x,t) = \frac{\gamma(t)}{x^{\alpha(t)}}  \ .
\end{equation}
Since Regge trajectories $\alpha(t)$ have positive slope, as
$-t $ increases they will reach some $-t\le -t_0$ where all
 $\alpha(t)$ become negative. There is partial evidence of this
from  Ref.~\cite{Coon:1974wh}.
At this point the amplitude becomes dominated by the contribution
by the $J=0$ pole.

 \subsection{ $J=0$ pole in Virtual Compton Scattering}
 \label{dvcs}

The main focus of deeply virtual Compton scattering  $\gamma^*(q) p
\to \gamma(q') p^\prime$  measured in $e p \to e^\prime p^\prime
\gamma X$ at large $q^2$ and the production of real photons,
$q'^2=0$ is the measurements of  the generalized parton
distributions (GPDs), which depend on both longitudinal and
transverse quark momenta.

However, DVCS also allows a new window into the study of the local
$J=0$ fixed-pole contribution which we have emphasized provides a
fundamental test of QCD.  Furthermore since the $J=0$ contribution
has a real phase, it has maximal overlap with Bethe Heitler
bremmstralung contribution to  $e p \to e^\prime p^\prime \gamma
X$~\cite{Brodsky:1971zh}.

As before we  work in the Bjorken limit, with $Q^2, \nu \to \infty$
with $x_B = Q^2/(2 p\cdot q)$  finite and $-t/Q^2 \to 0$
and write a fixed-$t$ dispersion relation in  $\nu$,
\begin{equation}
\nu = \frac{s - u}{4}   =  \frac{4 p \cdot q - Q^2}{4} = \frac{Q^2}{2\xi}
\end{equation}
We recall from Subsection.~\ref{leadingDVCS} that $\xi$ plays the role
of a generalized Bjorken variable for DVCS.
Using the $s-u$ crossing symmetry of $T_1$,  we have
\begin{equation}
T_1(Q^2,\nu,t) =  C_\infty(t) +
\frac{1}{\pi} \int_{\nu^2_{th}}^{\infty} \frac{d\nu'^2}
{\nu'^2 - \nu^2 - i\epsilon} \mbox{Im}T_1(Q^2,\nu',t) \ .
   \label{dispvcs}
\end{equation}
In analogy to doubly virtual Compton scattering we rewrite the
dispersion relation for $T_1$ at fixed $t$ in terms of dimensionless
variables $\xi$ and $x=Q^2/2\nu'$ in the scaling limit as \be
\label{dispDVCS} T_1(\xi,t)=C_\infty(t) +\frac{\xi^2}{\pi}
\int_0^1\frac{2dx}{x}\frac{{\rm Im}\ T_1(x,t)}{\xi^2-x^2-i\epsilon}
\ee Determining the subtraction constant, which  will turn out to be
the fixed pole contribution, requires additional information. If we
assume scaling and handbag dominance valid for $Q^2>>-t>-t_0$ with
momentum transfer such that Regge intercepts, $\alpha(t)<0$  are
negative,
\begin{equation}
\label{factorizedDVCS}
T_1(\xi,t)= - \int_{-1}^1 dx H(x,\xi,t) \left[ \frac{1}{x+\xi-i\epsilon} +\frac{1}{x-\xi+i\epsilon}
\right]\
\end{equation}
From it we can read that at high energies
\be \label{subtconstantDVCS}
C_\infty(t)=\lim_{\xi\to 0} T_1(\xi,t)= -2\int_{-1}^1 dx \frac{H(x,0,t)}{x}
 \ .
\ee
At large $-t$, $-t > -t_0$,  $H(x\to 0,0,t)\to 0$ and the integral is finite and real.
  At $\xi=0$ the $H(x,0,t)$ defines the generalized parton distribution functions discussed in Sec.~\ref{2VCS}
\begin{equation}
H(x,0,t) =\theta(x) f_q(x,t) - \theta(-x) {\bar f}_q(-x,t) \label{hvsf}
\end{equation}
with the two functions referring to the quark and antiquark
distributions, respectively. At $t=0$ they reduce to the parton
distribution functions measured in deep inelastic scattering. In
Sec.~\ref{2VCS} the single pdf $f(x,t)$ was used to denote the net
contribution from
 quark and antiquarks of given flavor. It is related to the two pdf's given above by
\begin{equation}
f(x,t) = f_q(x,t) + {\bar f}_q(x,t), \; x> 0  \label{ff}
\end{equation}
 We can also obtain the imaginary part of $T_1$ from
Eq.(\ref{factorizedDVCS})
to read
\be \label{imToffforward}
\frac{1}{\pi}{\rm Im\ }T_1(\xi,t)= H(\xi,\xi,t)-H(-\xi,\xi,t)
\ee
Substituting Eq.(\ref{subtconstantDVCS}) and Eq.(\ref{imToffforward}) in
Eq.(\ref{dispDVCS}) we obtain the analog of Eq.(\ref{T1forwardCS}),
\ba \label{DVCSlarget}
T_1(\xi,t)=-2\int_{-1}^1 dx \frac{H(x,0,t)}{x} \\ \nonumber
+ \xi^2 \int_{-1}^1 \frac{dx}{x} \frac{H(x,x,t)-H(-x,x,t)}{\xi^2-x^2-i\epsilon}
\ea
We can also write $T_1$ as
\begin{eqnarray}
 \label{DVCSlarget-as}
T_1(\xi,t) &= &    \int_{-1}^1  dx \frac{x}{\xi^2 - x^2 -i\epsilon}H^+(x,x,t)   \nonumber \\
 &+ & \int_{-1}^{1} \frac{dx}{x}  \left[ H^+(x,x,t)   - H^+(x,0,t) \right]  \nonumber \\
\end{eqnarray}
where we defined the positive charge conjugation, $H^+$ generalized parton distributions,
\begin{equation}
H^+(x,x,t) \equiv H(x,x,t) - H(-x,x,t)
\end{equation}
and
\begin{eqnarray}
H^+(x,0,t) & = &   H(x,0,t) - H(-x,0,t) \nonumber \\
& =& \theta(x) f(x,t) - \theta(-x) f(-x,t)
\end{eqnarray}
with $f(x,t)$ given by Eq.(\ref{ff}).
As long as $-t > -t_0$ with all Regge intercepts
negative $\alpha(t) < 0$, $H^+(x,x,t)$ vanishes in the limit $x\to0$ and  the $J=0$
pole contribution to $T_1$  is given by
\begin{eqnarray}
T_1^{J=0}(t) & = &  \lim_{\xi \to 0} T_1(\xi,t)  =  -\int_{-1}^{1} \frac{dx}{x} H^+(x,0,t) \nonumber \\
&  = &  -2 \int_{-1}^{1} \frac{dx}{x} H(x,0,t) \label{T1DVCSV}
 \end{eqnarray}
Now we are ready to lift the assumption that $-t> t_0$ and
 consider the situation where  some of the intercepts are positive,  $\alpha(t) >0$, still assuming Bjorken scaling.
Since $T_1$ is finite for any $\xi$ it follows from Eq.(\ref{DVCSlarget-as}) that
\begin{equation}
\lim_{x \to 0} \left[ H^+(x,x,t) -  H^+(x,0,t) \right] = 0  \label{tmp}
\end{equation}
Thus we define the valence part $H_v$  as the part of $H$ which is finite in the $x\to 0$ limit,
\begin{equation}
 \label{Hvalence-x}
H_v(x,x,t) \equiv  H(x,x,t) - H_R(x,t)
\end{equation}
which implies  $H^+_v(x,x,t) \equiv H_v(x,x,t) - H_v(-x,x,t)$ and similarly, following Eq.(\ref{tmp})
\begin{equation}
 \label{Hvalence}
H_v(x,0,t) \equiv  H(x,0,t) - H_R(x,t)
\end{equation}
and  $H^+_v(x,0,t) \equiv H_v(x,t) - H_v(-x,t)$ which all finite in
the limit $x\to 0$. Here we assumed that in general the small-$x$
behavior is
 of the Regge type given by,
\begin{eqnarray}
& & H_R(x,t) \equiv \theta(x) \sum_{\alpha>0}\frac{\gamma_\alpha(t)}{x^{\alpha(t)}} - \theta(-x)\sum_{\bar \alpha>0}
\frac{\bar\gamma_{\bar \alpha}(t)}{(-x)^{\bar \alpha(t)}}  \nonumber \\
& & H_R^+(x,t) \equiv H_R(x,t) - H_R(-x,t)
\end{eqnarray}
Finally for the amplitude  $T_1$ we obtain,
\begin{widetext}
\begin{equation}
 \label{DVCSlarget-as2}
 T_1(\xi,t)=     \int_{-1}^1  dx \frac{x}{\xi^2 - x^2 -i\epsilon}H^+_v(x,x,t)
 +   \int_{-1}^1  dx \frac{x}{\xi^2 - x^2 -i\epsilon}H^+_R(x,t)
+ \int_{-1}^{1} \frac{dx}{x}  \left[ H^+_v(x,x,t)   - H^+_v(x,0,t) \right]
\end{equation}
\end{widetext}
To extract the $J=0$ pole contribution we need to study the $\xi\to 0$ limit.  In this limit we find
\begin{eqnarray}
\lim_{\xi \to 0} T_1(\xi,t)
& = &   - \int_{-1}^1\frac{dx}{x} H^+_v(x,0,t)  \nonumber \\
& + &  \lim_{\xi \to 0} \int_{-1}^{1} dx \frac{x}{\xi^2 - x^2 - i\epsilon} H^+_R(x,t)
\label{subtracted-1}
\end{eqnarray}
The second term on the {\it r.h.s} in Eq.(\ref{subtracted-1}) yields
terms with  grow with energy as $\sim 1/\xi^{\alpha(t)}$,
$(\alpha(t) > 0)$, a constant term  $O(\xi^0)$,   and sub-leading
terms which decrease with increasing energy, $O(\xi^{2-\alpha(t)})$.
Extracting the constant term from the Regge term we obtain for the
$J=0$ contribution to $T_1$,
\begin{eqnarray}
T_1^{J=0}(t) & = &  -2\int_{-1}^{1} \frac{dx}{x} H_v(x,0,t)  \nonumber \\
&  + &  2\sum_{\alpha >0} \frac{\gamma_\alpha(t)}{\alpha(t)}
 + 2\sum_{\bar \alpha>0} \frac{\bar \gamma_{\bar \alpha(t)}}{\bar \alpha(t)}\label{subtracted}
\end{eqnarray}
We note that since $H_v(x,0,t)$ is the generalized parton distribution,
the expression for the $J=0$ pole contribution in DVCS with real photon in the final state
is exactly the same as obtained in Sec.~\ref{2VCS} for doubly virtual Compton scattering.

We have thus shown that the $J=0$ pole contribution is universal,
{\it i.e} the same for doubly virtual Compton scattering, as given
by Eq.(\ref{2vcsfinal2}) and in  virtual Compton scattering, as
given by Eq.(\ref{subtracted}). Again, as the momentum transfer
increases, the intercepts $\alpha$ become negative and the virtual
Compton amplitude at high energies, $\xi \to 0$ becomes dominated by
the $J=0$ pole contribution, while other details of nucleon
structure described by the valence generalized  parton distribution,
$H_v(x,\xi,t)$ are suppressed by powers of $\xi^{1-\alpha(t)}$ where
$\alpha(t)<0$ is the largest  intercept.

Bjorken scaling in DVCS  has been demonstrated to hold \cite{Collins:1996fb,Collins:1998be} from an analysis of QCD
corrections to the elementary two-photon parton amplitudes. In
particular it has been shown that  IR divergences can be absorbed
into the soft parton-nucleon amplitudes parameterized by the
generalized
 parton distributions. This proof however relies on the assumption
that GPD's are finite at the break points, {\it i.e.} that  the limit
$\lim_{x \to \xi} H(x,\xi) $ exists.  If  $ \mbox{ Im} T_1(\nu,Q^2)$ does not scale {\it i.e.}
$ \mbox{ Im} T_1(\nu,Q^2) \sim \nu^{\alpha(t)}$ with $\alpha(t)>0$ at
small-$t$ then in the Bjorken limit $ \mbox{ Im} T_1(\nu,Q^2)$ diverges
 and so does  $\lim_{\xi \to \xi} H^+(\xi,\xi)  = \infty $. The converse is also true,
in such a case one would expect
    $T_1(\nu,Q^2) \sim (Q^2/\xi)^{\alpha(t)}$.
A possibility of non-scaling  in DVCS and its consequences for the
high energy behavior has been studied in Ref.~\cite{Szczepaniak:2007af}
and shown to be consistent with scaling observed in  doubly virtual
Compton amplitude.  We note, however, that the $J=0$ contribution, being $\nu$-independent
is truly universal, regardless of the  scaling properties of $T_1$.

\subsection{Real Compton Scattering (RCS) }\label{sec:rcs}

The first evidence for the existence of a $J=0$ contribution in the real Compton
amplitude was developed by Damashek and Gilman, based on the dispersion theory and
measurements of the total photoabsorption cross section $\sigma(\gamma p \to X).$

The Gell-Mann, Goldberger, Thirring subtracted dispersion relation for the forward
helicity no-flip Compton amplitude $T_1(\nu)= T_1(\nu,Q^2=0)$ is
\begin{equation}
T_1(\nu) = T_1(0) + {\nu^2\over \pi}  \int^\infty_{\nu_0}{d{\nu^\prime}^2\over {\nu^\prime}^2-\nu^2 - i\epsilon}
{{\rm Im} T_1(\nu^\prime)\over{ \nu^\prime}^2} \label{rcs}
\end{equation}
where $\nu = (s-u)/4$
and $\nu_0$ corresponds to the pion production threshold, $\nu_0 = m_\pi M_N + m_\pi^2/2$.

As noted earlier,  the low energy, $\nu \to 0$, the  Compton
amplitude has local and non-local contributions  ({\it cf.}
Eqs.(\ref{NL}),(\ref{L}))
\begin{equation}
T_1(0)  = T^L_1 + T^{NL}_1.
\end{equation}
and $T_1(0)=-2$ is the Thomson term.  Note that $T^{NL}_1$,
includes the contributions of the cat's ears diagrams where the
incident photon couples to one quark current and the outgoing photon
couples to the other current.
The local term, $T^L_1$  is real;
only the non-local contribution has an absorptive, imaginary part.
 For RCS we can define $x \equiv \nu_0/\nu$ so the amplitude parallels that of virtual Compton scattering.
\begin{equation}
T_1(x) =   T_1(0)
 +   \frac{1}{\pi}\int_0^1 dx' \frac{2x'}{x^2 - x'^2 - i\epsilon} \mbox{ Im}T_1(x')
\end{equation}
Subtracting the Regge contribution and defining,
\begin{equation}
f_v(x) \equiv \frac{1}{\pi} \mbox{ Im}T_1(x) - \sum_{\alpha \ge 0}  \frac{\gamma_\alpha}{x^\alpha} \nonumber \\
\end{equation}
gives for the energy independent, $J=0$ contribution,
\begin{equation}
T^{J=0}_1 =  T_1(0) -2 \int_0^1 \frac{dx}{x} f_v(x)  +  2\sum_{\alpha>0} \frac{\gamma_\alpha}{\alpha}
\end{equation}
At high energy, $\nu \to \infty$, $x\to 0$, the quark-photon
interactions are universal, and we thus expect the $J=0$
contribution to be identical to the local, $T^L_1$ term,
\begin{equation}
T^{J=0}_1 = T^L_1
\end{equation}
In the limit $t\to 0$ this relates the normalization of the $1/x$ moment of
the forward RCS amplitude to that of the DVCS in the limit $-t \to 0$,
\begin{equation}
 \int_0^1 \frac{dx}{x} f_v(x,Q^2 \to \infty, t=0)   =   1
  +   \int_0^1 \frac{dx}{x} f_v(x,Q^2 = 0,t=0)
\end{equation}
For finite momentum transfer $t\ne 0$ the constant  on the {\it
r.h.s} should be replaced by $-1/2$ of the finite-t subtraction
constant from Eq.(\ref{rcs}), which for fixed angle scattering is
expected to fall off as a power of $-t$, and one  would expect  the
$1/x$ moment in DVCS and RCS be identical at large $s$ and large
$-t$, with $-t/s\sim O(1)$
\begin{equation}
\int_0^1 \frac{dx}{x} f_v(x,Q^2 \to \infty, t ) =  \int_0^1 \frac{dx}{x} f_v(x,0,t)
\end{equation}
Damashek and Gilman have used the dispersion relation in
Eq.(\ref{rcs}) and the measured photoabsorption cross section to
determine  ${\rm Re}T_1(\nu).$ They fit the high energy
photoabsorption cross section to $s$-channel resonances at low
energies and the Pomeron $\alpha_P(0)=1$ and Reggeon
$\alpha_R(0)=1/2$ contributions at high energies: $\sigma= \sum_i
c_i \nu^{\alpha_i(0) -1}$.  Since the Pomeron contribution has an
imaginary phase, this form predicts a Reggeon $\nu^{\alpha_i(0)} =
\nu^{1/2}$ contribution at high energy for the forward amplitude,
which as we have argued is associated specifically with
$T^{NL}_1(\nu)$. However, Damashek and Gilman also find that the
dispersion relation predicts an additional constant contribution to
${\rm Re}T_1(\nu)$  at high energies. As we discussed above this can
be identified with the   local term  $T^L_1.$
   Since this constant term is found empirically to have approximately the same
   value as the Born term, this implies
   \begin{equation}
   -2 \int_0^1 \frac{dx}{x} f_v(x) + 2\sum_{\alpha >0} \frac{\gamma_\alpha}{\alpha} \sim 0
   \end{equation}
   and that the $J=0$ fixed pole on the proton at $t=0$  has the
   value $-2 \langle \sum_q e^2_q / x_q \rangle  \sim -2 \langle \sum_q e_q \rangle^2 =-2$
     in the energy domain of the photoabsorption experiment.
An interesting test of this analysis would be the measurement of the photoabsorption
cross section on a neutron target.
  One predicts a fixed pole a factor of  2/3
 smaller (see Eq.(\ref{npratio}) below).

\section{ Phenomenology of the $J=0$ pole } \label{sec:forward}
\subsection{The $J=0$ pole in the forward limit and the structure function parametrization}  \label{subsec:pdfs}

The $J=0$ pole in the forward limit $t=0$ is given
in terms of parton distribution functions measurable in DIS.
Parton distribution functions need to diverge at small $x$ due to the
Regge behavior of hadron scattering amplitudes and simple statistical
arguments.
This well known observation, supported by the extensive HERA data at small
$x$, was made by Kuti and Weisskopf \cite{Kuti:1971ph}.
The Lorentz invariant phase space of a parton whose
transverse momentum can be ignored is $dp_i/2E_i=dx/2x$. If one ignores
dynamical effects, and imposes the statistical hypothesis that all
states
for the sea partons are equally likely, one sees that the sea
distribution
functions scale as
\begin{equation}
\lim_{x\to 0} f_s(x) = \frac{\gamma_1}{x}.
\end{equation}
The small $x$ divergence in the sea part of the pdf reflects simply large
phase space available at large energy. Kuti and  Weisskopf further
observed that the Regge behavior of the photon-proton scattering
amplitude needs to stem from Regge behavior of the parton distribution
functions within the proton, and found that the small $x$ behavior if pdf's is given by
\begin{equation}
\lim_{x\to 0} f(x) =  \frac{\gamma_\alpha}{x^{\alpha(0)}}
\end{equation}
with $\alpha(0)$ the usual intercept of the Regge trajectory with the
$t=0$ axis in a Chew-Frautschi plot.  As discussed in previous section the pdf's with $\alpha<0$
are to be associated with the valence distributions, and $\alpha >0$ with the sea distribution.
The split between sea and valence Regge contribution
supports the interpretation that point-like current interactions  on
target constituents is dual to exchanges of all
residual, or daughter   Regge trajectories.

Modern fits to deep inelastic scattering data routinely employ
a small $Q^2$ parameterization of the pdf's which  is a simple variation of
the Kuti-Weisskopf statistical model, namely \cite{Martin:1998sq}
(see table I in that paper for the parameters $A_i$, $\eta_i$, $\lambda_i$
$\epsilon_i$, $\gamma_i$)
 \begin{equation} \label{KWfits}
xf_i(x) = A_i x^{\eta_i} (1-x)^{\lambda_i}(1+\epsilon_i \sqrt{x}
+ \gamma_i x).
\end{equation}
We expose its Regge form around $x=0$ by expanding the $(1-x)$ in powers of $x$,
\ba
f_i(x) =  A\left(x^{\eta_i-1} +\epsilon_i x^{\eta_i-1/2}
+(\gamma_i-\lambda_i) x^{\eta_i} \right. \\ \nonumber \left.
-\epsilon_i\lambda_i x^{\eta_i+1/2} -\gamma\lambda x^{\eta_i+1}
+\dots
\right)
\ea
In terms of these MRST  \cite{Martin:1998sq}  parameters, the Regge intercept is
 $\alpha(0)=1-\eta_i$.
Phenomenology of deep inelastic
scattering requires $\eta$ to be smaller than 1 for several pdf's.
For the valence flavors, a typical Regge intercept is of order
$\alpha(0)=1/2$. This is the case for the GRV98 pdf set \cite{Gluck:1998xa}
that has exponents $\alpha(0)=-0.85$ and $-0.52$ for the light sea and valence pdf's respectively.
(Here ``valence'' is used in the sense
of Ref.~\cite{Martin:1998sq}.)
Notice that the $\sqrt{x}$ in Eq.(\ref{KWfits}) gives
rise to subleading Regge power laws.
For the MRST98  \cite{Martin:1998sq}  pdf sets, an also widely used
alternative, the power law exponents have higher variation around
classical Regge theory and the $u$ proton's valence component has a
somewhat high intercept $\alpha_u(0)\in(0.53,0.59)$, the $d$ valence
component being definitely at odds with other phenomenology with
$\alpha_d(0)\simeq 0.73$ as large as the sea component. The subleading
Regge behavior is also  given by the $\sqrt{x}$ factor in Eq.(\ref{KWfits}),
and having an  intercept larger than zero, it also causes
a divergence. In both GRV98, MRST98 sets the gluon pdf behaves as a
valence-like parton with a very small intercept at this low scale,
indication of the gluon degrees of freedom being gapped at low energy \cite{Alkofer:2006xz}.

Since $T_1^{J=0}$ was initially assessed in
Ref.~\cite{Brodsky:1973hm} there has been immense experimental
progress and accurate pdf fits for a wide range of $Q^2$ are now
available. We have been conservative and chosen a decade-old set of
fit functions which have been thoroughly tested.

\begin{table}
\begin{tabular}{|c|ccc|cc|}
 \hline
quark  & MRST  &MRST  &MRST & LO  &NLO\\
 flavor & Low gluon & Central gluon &Upper gluon  & GRV & GRV\\
\hline
$u$       & $51$  & $14$ & $13$   & $-36$ & $17$ \\
$\bar{u}$ &$-5.3$ & $-1.3$ & $-7.0$ & $62$  & $9.7$\\
\hline
$d$       & $6.1$ & $5.9$ & $5.0$ & $-120$ & $-11$\\
$\bar{d}$ & $-0.78$  & $-0.46$  &$-1.8$  & $-62$  & $-13$ \\
\hline
$s$       & $-1.5$ & $-0.43$ & $-2.2$ & 0 & 0\\
\hline
$\sum$ & $50$ & $18$ & $7.0$ & $-160$ &$2.7$  \\
\hline
\hline
\end{tabular}
\caption{
 $F^{C=+}(0)$ defined in Eq.(\ref{deff}), that is
$
-\frac{1}{2} T_1^{J=0} = \int_0^1 \frac{dx}{x} (f(x)-f^R(x)) -
\sum_{\alpha>0} \frac{\gamma_\alpha}{\alpha}
$
for MRST98 \cite{Martin:1998sq} and GRV \cite{Gluck:1998xa} full
parton distribution functions.
 We take both sets at the low scale defining the fit
parameters since we prefer analytical expressions to better control the
subtractions. These scales are $1\mbox{ GeV}^2$ for the MRST set
and $0.26(0.4) \mbox{ GeV}^2$ for the LO(NLO) GRV set. The latter has no strange sea component at
this low scale.
A large spread in the results  comes from the uncertainty in the  subtraction constants
  $\gamma/\alpha$, which are  not yet very well determined.
\label{tableweisberger}}
\end{table}

In Table \ref{tableweisberger} we have collected the values of the
integrals in Eq.(\ref{Weisberger})
multiplied by the corresponding quark to positron charge ratios $e_q^2$. We have explicitly separated the
contribution from quark and antiquark, and subtracted all Regge poles with
$\alpha>0$  from $f$, $\bar{f}$ to yield $f_v$ and
$\bar{f}_v$ (which are finite at $x=0$). This is also equal to Eq.(\ref{2vcsfinal2}) up to a (-1/2) factor.

The sum over the entries in a column in this table should be a
number of $O(1)$, judging by the results of
Ref.~\cite{Damashek:1969xj}. This relies on a cancelation which does
not occur using the  current parameterizations  of pdfs. This
possibly signals a systematic uncertainty in the way these
parameterizations are written down. As can be seen the results are
spread over an order of magnitude, and in some cases even the sign
is difficult to ascertain. The reason is simple. The regulated
integral over the valence pdf's is perfectly well-behaved. However,
subleading Regge poles with an intercept just below zero, which are
integrable and need no subtraction, will yield a non-negligible
contribution. The MRST and even the GRV fit which seem to comply
better with the theoretical arguments of Kuti and Weisskopf, employ
a formula such as Eq.(\ref{KWfits}). There, the subleading terms
with $\sqrt{x}$ and $x$ are phenomenologically added to improve the
fit to structure function data. For most applications the precise
power law at small $x$ of $f$ is not needed, a computer code which
yields $f$ as usually provided, suffices, but to have the best
possible fits becomes critical if we enhance the small $x$ part by
computing the $1/x$ moment. This can be seen by considering the
following toy distribution function
\begin{equation}
f(x)=x^\alpha (1-x) (1+\alpha)(2+\alpha) \ .
\end{equation}
The expectation value of the momentum fraction for the parton is
\begin{equation}
\langle x \rangle = \frac{1+\alpha}{3+ \alpha}
\end{equation}
which converges to $1/3$ in the $\alpha \to 0$ limit, but then
\begin{equation}
\left\langle \frac{1}{x} \right\rangle = \left(1+\frac{2}{\alpha}\right)
\end{equation}
which is divergent in the same limit. Therefore we see how it is
really critical to control Regge poles with intercept near and below
zero. Indeed this is not the case for the standard pdf
parameterizations, and while a group
   of trajectories bunch around intercept values 0.4 to 0.6, others are well below zero.

Another  comment is in order:  the recent G0 and Happex experiments at
Jefferson Lab \cite{Armstrong:2005hs} have not yet settled what level of
strange sea is needed to account for parity violation even at small $Q^2$.
Therefore one should take the GRV fit where the strange
sea vanishes with caution.
Another difference between the MRST and the GRV fits worth recalling is that
the isospin asymmetry disappears in the Regge limit of the simplest MRST
set we employ, not
so in the GRV fits where it is controlled by a standard Reggeon of intercept
$\simeq 1/2$.
Although not much can be said from Table \ref{tableweisberger}, it
seems
that the integral over the valence pdf's is negative.

\subsection{The $1/x$ moment and Weisberger's relation.}
\label{Weisbergersection}
The importance of the $1/x$ moment of parton distribution functions,
which measures the valence quark contribution to the $T_1^{J=0}$
amplitude  was stressed by Weisberger in
Ref.~\cite{Weisberger:1972hk}.  There he derived a relation between
this $1/x$ moment and the derivative of the squared proton mass with
respect to the squared parton mass taking  into account only the
kinetic energy dependence on the quark mass. The resulting relation
is valid for the parton mass and the corresponding distribution
function defined at the same scale $\mu$. In modern notation and
normalization \cite{Martin:1998sq}, Weisberger's result reads
\begin{equation} \label{Weisberger}
\frac{\delta M_N^2}{\delta m_i^2(\mu)} = \frac{1}{e_i^2}\int_0^1 \frac{dx}{x}
[f_{v,i}(x)_\mu+\bar{f}_{v,i}(x)_\mu].
\end{equation}

Here $f_{v,i}/e_q^2$ is the $i$th-valence quark distribution function,
and likewise for the antiquark (once the Regge part has been subtracted).
  This relation can be easily
understood following the formal argument of Weisberger. He observed that
the shift of the nucleon's energy $E$ upon shifting
the energy of a parton $E_i$ in the Bjorken limit is given by
\begin{equation} \label{energyshift}
\delta E= \frac{1}{e_i^2}\int_0^1 dx_i f_{v,i}(x_i) \delta E_i(x_i) \ .
\end{equation}
In this limit parton momentum is taken to be parallel  to that of the
nucleon,
$p_i=x_i P$ and for the parton with mass $m_i$ one has,
$$E_i^2 = m_i^2 + x_i^2 P^2$$
 so that
$$ 2E_i \delta E_i = \delta (m_i^2)\ \   {\rm or}\ \
 \delta E_i= \frac{\delta (m_i^2)}{2x_i P} \ .$$
Identifying $E$ with nucleon mass
and using $\delta M^2 = 2 E \delta E$ we find
\begin{equation}
\delta M^2 = \delta E^2 = \frac{1}{e_i^2}\int_0^1 dx_i f_{v,i}(x_i) \delta m_i^2  \frac{2E}{2Px_i}
\end{equation}
from which Eq.(\ref{Weisberger}) follows. On first impression one
would think of taking also a derivative of the pdf respect to the
quark mass, but this is not the case  according to the
Hellman-Feynman theorem.


One can see that Weisberger's result holds simply by noting that in
light front quantization the Hamiltonian contains a kinetic energy
term
\begin{equation}
M^2_{\rm kin}=\sum_i \frac{k^2_\perp + m^2_i}{x_i}
\end{equation}
and no other explicit quark mass dependence.
Upon taking the expectation value $\langle \delta M^2/ \delta m_i^2
\rangle$ in the nucleon state  and, in the collinear approximation,  ignoring the $k_\perp$
 immediately  leads to  Weisberger   relation of Eq.(\ref{Weisberger}).
 In light-front QCD there is one further implicit quark mass
dependence
in the quark-gluon vertex. This is analogous to
the QED case, where the spin-flip vertex term $e \to e \gamma$ in the
QED LF Hamiltonian is proportional to $m_e$
 ({\it e.g.} Table 6 on page 78 of  Ref.~\cite{Brodsky:1997de}).
This yields an additional  contribution to the Weisberger relation.

The Weisberger relation involves the proton state, which by
definition is  normalizable,  and therefore contains only bound
constituents; i.e., any contribution to the structure functions
which can be interpreted as originating from processes in which
photon splits into the $q\bar q$ pair which later re-scatter off
proton' constituents should not be included in
Eq.(\ref{Weisberger}). Of course since we do not know proton's wave
function and pdf's are known through fits to data  rather than from
first principle calculations, the separation of the types of
processes is at best phenomenological. Since the left hand side in
the Weisberger relation is finite, the valence structure function
entering the right hand side has to satisfy $\lim_{x\to0} f_{v,i}(x)
= 0 $. We thus take it to be given by Eq.(\ref{valence}), and the
values of $\delta M^2_N/\delta m_i^2$ obtained for different pdf
parameterizations are given in Table~\ref{tableweisberger}.


\begin{table}
\begin{tabular}{|c|ccc|cc|}
 \hline
quark  & MRST  &MRST  &MRST &
LO  &NLO\\
 flavor & Low gluon & Central gluon &Upper gluon  &
GRV & GRV\\
\hline
$u$       & -6.7  & -7.0 & -11   & 12 & 10.6\\
$\bar{u}$ &-20    &-16   & -20 & -12  & -12\\
$\frac{\delta M_N^2}{\delta m_u^2}$ & -27 & -23 & -31 & $\simeq 0$ & -1.1\\
\hline
$d$       & -39 & -52 & -39 & 120 & 130\\
$\bar{d}$ &-27  &-27  & -33  & 67  & 70\\
$\frac{\delta M_N^2}{\delta m_d^2}$ & -66 & -79 & -72 & 180 & 200\\
\hline
$s$       & -15 & -22 & -29 & 0 & 0\\
$\frac{\delta M_N^2}{\delta m_s^2}$ & -31 & -45 & -58 & 0 & 0\\
\hline
$g$       & $\simeq 600$ & $\simeq -350$ & $\simeq 1500$ & 4.4 & 12\\
\hline
\end{tabular}
\caption{ As in table \ref{tableweisberger}  but without the
$\sum_{\alpha>0} \frac{\gamma_\alpha}{\alpha}$ terms, that is,
corresponding to Eq.(\ref{Weisberger}). Again systematic differences
appear between the MRST and GRV sets, but the different MRST
parameterizations are now very consistent.
 \label{tableweisberger2}}
\end{table}

\section{ The $1/x$ form factor and off-forward $J=0$ pole }
\label{sec:simplemodels}

In the previous section we considered the $J=0$ component of the $T_1$
amplitude in the forward limit of doubly virtual Compton scattering
and its relation to the Weisberger sum rule. Now we return to finite momentum transfer,
where the $t$-dependence of the fixed pole
$C_\infty(t)$ provides a new form factor of the nucleon.

In analogy with the conventional Dirac form factor expressed in terms of
the Generalized Parton Distribution
\be
F(t)= \int_{-1}^1 dx H(x,0,t)
\ee
At sizable $-t$ when Regge intercepts are negative, the $1/x$ form factor defined in
Eq.(\ref{deff}) is given by ({\it cf.}
 Eq.(\ref{T1DVCSV}))
\begin{equation}
F_{1/x}(t) =   \int_{-1}^1 \frac{dx}{x} H(x,0,t) \label{fused}
\end{equation}
and for small $t$ where there can be Reggeons with intercept
$\alpha >  0$ according to Eq.(\ref{subtracted}) we find,
\ba
F_{1/x}(t) \equiv  \int_{-1}^1 \frac{dx}{x} H_v(x,0,t) -
\sum_{\alpha>  0} \frac{\gamma_\alpha}{\alpha(t)} -
\sum_{\bar \alpha>  0} \frac{ \bar \gamma_{\bar \alpha}}{\bar \alpha(t)}\\ \nonumber
\ea
As momentum transfer increases in virtual  Compton scattering the
energy dependent part of  the transverse photon amplitude decreases
at high energies and $F_{1/x}(t)$ is expected to dominate the cross section.
As discussed in Section.~\ref{dvcs} the $J=0$ pole contribution to the
 transverse amplitude in singly virtual Compton scattering is given by
the $J=0$ component of the doubly virtual amplitude.

There has been much progress in the extraction of the conventional
Dirac form factor using lattice QCD~\cite{sachrajda} thanks to the
use of twisted boundary conditions; in principle the $F_{1/x}$ can
also be extracted. Meanwhile we can provide an estimate using
 models of generalized parton distributions.
We restrict ourselves to three simple models:  ansatz in terms of
double
distributions~\cite{Radyushkin:1998es,Polyakov:1999gs,Lehmann-Dronke:2000xq,Anikin:2004jb},
the light-front wavefunction representation in the valence
constituent quark model of the proton~\cite{Brodsky:2000xy}, and the
quark-diquark model~\cite{Brodsky:2005vt}. We comment on how the
$J=0$ fixed pole behavior is universal and arises for any reasonable
model of the GPD's -- as it should.

\subsection{Double distribution parametrization of GPD's}
 \label{Radyushkin}

  In this ansatz GPD's are computed in the symmetric frame through
\begin{widetext}
\begin{eqnarray} \label{anikins}
H^q(x,\xi,t)=   \frac{F_1^q(t)}{F_1^q(0)} \left(
\frac{\theta(\xi+x)}{1+\xi}
\int_{0}^{{\rm min}\{ \frac{x+\xi}{2\xi},
\frac{1-x}{1-\xi} \}} dy\,
F^q(x_+, y) -
\frac{\theta(\xi-x)}{1+\xi}
\int_{0}^{{\rm min}\{ \frac{\xi-x}{2\xi},
\frac{1+x}{1-\xi} \}} dy\,
F^{\bar q}(x_-, y) \right)
\\ \nonumber
+\theta(\xi-\arrowvert x \arrowvert) D(x/\xi,t)/N_f
\ .
\end{eqnarray}
\end{widetext}
In the original parametrization without $D$-term, $t$ dependence as
factorized and parameterized in terms of the proton's $F_1$ form
factor,  which is usually taken as a  dipole
\begin{equation}
F_1(t)  = \frac{1}{\left( 1 - \frac{t}{0.7 \mbox{ GeV}^2}\right)^2}
\left[1 - \frac{t \mu }{4M_N^2}\right]
\end{equation}
with $\mu \simeq 2.793$ being the anomalous magnetic moment.
The $x_\pm$ variables in  Eq.(\ref{anikins}) are defined by,
\begin{equation}
x_+ = \frac{x+\xi-2\xi y}{1+\xi},  \
x_- = \frac{\xi-x -2\xi y}{1+\xi} \ .
\end{equation}
 Without the $D$-term, the momentum-transfer dependence of the
$1/x$ will thus be the same as $F_1(t)$
\begin{eqnarray}
F^q_{1/x}(t)  & =  &  \int_{-1}^1 {{dx} \over {x}}  H(x,\xi=0,t)  \nonumber \\
& = &  F_1(t) \int_{-1}^1 {{dx} \over {x}}  H(x,\xi=0,0) \label{fac}
\end{eqnarray}
We note  that this  will be the same in any factorizable ansatz for the
Generalized Parton Distributions.  In the limit $\xi \to 0$ the
first and second term on the right hand side of Eq.(\ref{anikins})
contribute  to Eq.(\ref{fac}).
The last term, in Eq.\ref{anikins},  the so called  $D$-term,
\cite{Polyakov:1999gs}, gives a contribution to $H$ which depends on
the fraction $y\equiv x/\xi. $ It leads to a contribution to the
DVCS amplitude which is part of the fixed pole
 and remains finite and real  in the $s \to \infty$  ($\xi \to 0$)
limit.  Its contribution to the $F_{1/x}$ form-factor
given by
\begin{equation}
F^D_{1/x}(t) = F^D_{1/x}(0) =  {1\over {N_f}}
\int_{-1}^1\frac{dy}{y} D(y,t).
\end{equation}
It worth nothing that the fixed-pole receives contribution from the valence part of the
quark double distribution, as well as the $D$-term.

\subsection{Light-front constituent quark wavefunctions}
\label{subsec:lightcone}

Following Ref.~\cite{Brodsky:2000xy} we write the light-front
valence constituent quark model representation of the GPD. If higher
Fock space components of the proton are suppressed, the
only interval where $H(x, \xi;t)$ is non-vanishing is
$\xi < x < 1$, whereas the particle-number changing contribution in
the interval $-\xi< x< \xi$ and the antiparticle contribution in
$-1 < x < -\xi$ are absent.

\begin{figure}
\includegraphics[height=2.5in]{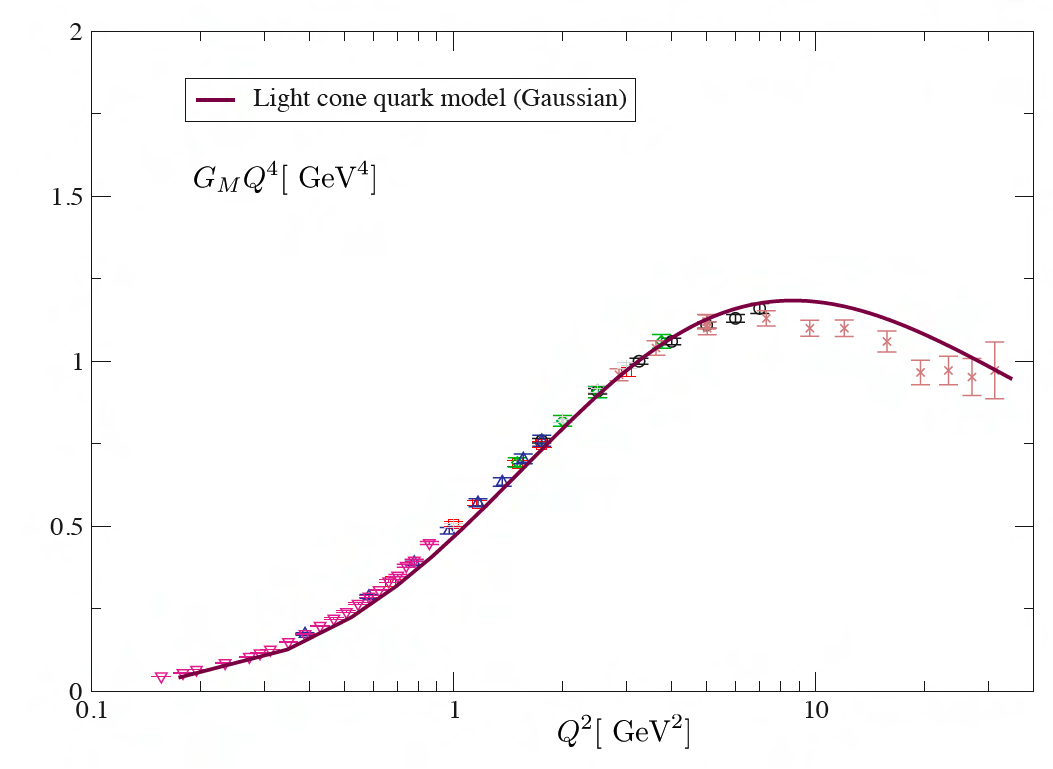}
\caption{\label{formfactorplot}
Proton Sachs form factor in the impulse approximation with light-front
Gaussian wave function of Eq.(\ref{lcwf})  and one-gluon exchange.
  The parameters employed are $m=250\ MeV$ and
 $b=1.2\mbox{ GeV}^{-2}$.  Data are taken from  Ref.~\cite{Brash:2001qq}.}
\end{figure}

\begin{figure}
\includegraphics[height=2.3in]{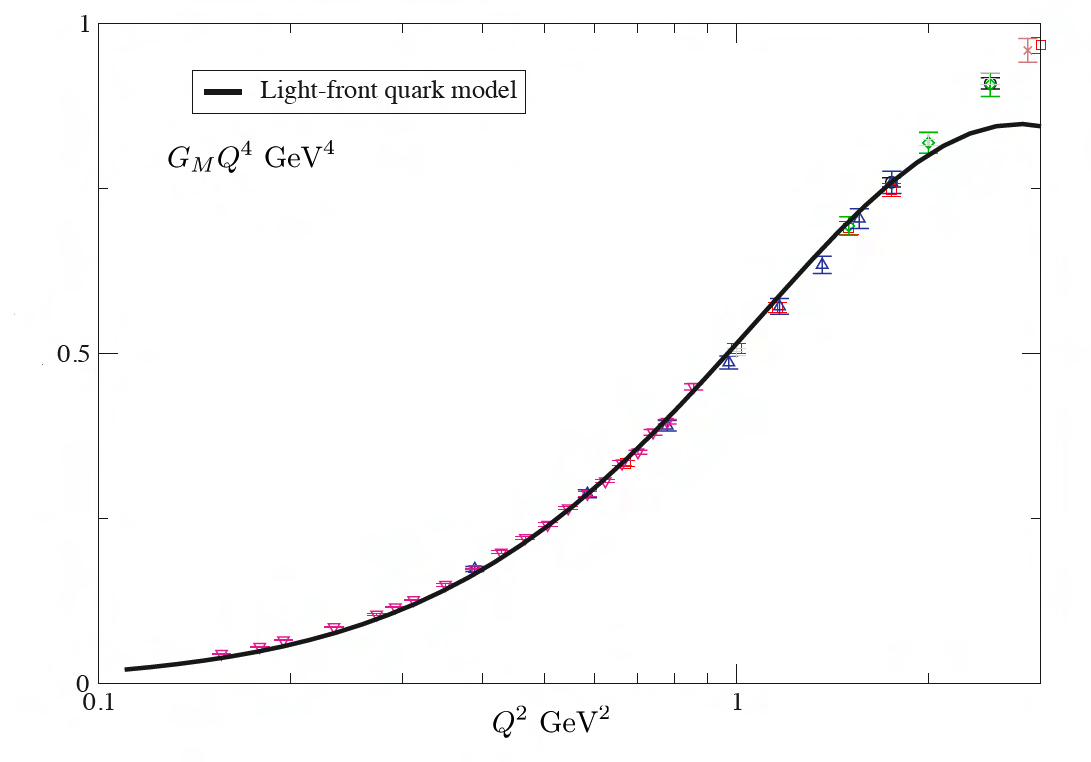}
\caption{\label{fig:softformfactor}
Proton Sachs form factor in the impulse approximation with light-front Gaussian wave function from Eq.(\ref{lcwf}).  Since this approximation excludes the hard, one-gluon exchange component
  (as opposed to Fig.~\ref{formfactorplot}), the form factor
representation can only be trusted until about $3\mbox{ GeV}^2$.}
\end{figure}

\begin{figure}
\includegraphics[height=2.3in]{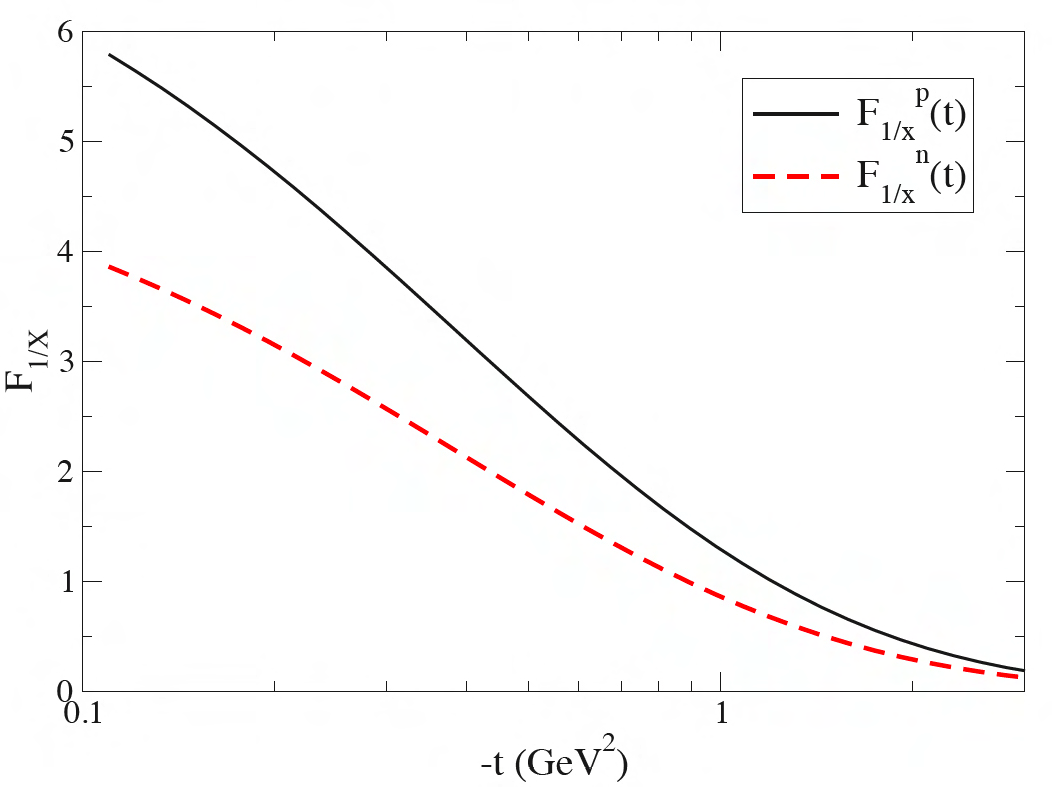}
\caption{\label{fig:1overxlightfront}
The $1/x$ form factor of the proton and neutron in the valence light-front
quark model. The ratio is equal to  $2/3$  arising from the ratio of quark electric charges.
Parameters are as in Fig.~\ref{formfactorplot}}
\end{figure}

\begin{widetext}
\begin{eqnarray}
H(x,\xi,t) &= &  \frac{\theta(x-\xi)}{(1-\xi)}
\sum_{\lambda_i}\int \prod_{i=1}^3 \left(\frac{dx_i d^2{\bf k}_{\perp\ i}}
{16\pi^3}\right)  16\pi^3 \delta(1-x_1-x_2-x_3) \delta^{(2)}
(\sum {\bf k}_{\perp\ i})  \nonumber \\
& \times &  \delta(x-x_1) \psi_{\lambda'=+}^*(x'_i,{\bf k}'_{\perp\
i},\lambda'_i) \psi_{\lambda=+}(x_i,{\bf k}_{\perp i},\lambda_i).
\end{eqnarray}
\end{widetext}
Here the light-front quark momentum fractions in the final proton
 are $x'_i= (x_i(1+\xi)-2\xi)/(1-\xi)$  for the struck quark and
   $x'_i = x_i(1+\xi)/(1-\xi)$ for spectators, respectively.
   A simple Gaussian model is commonly used \cite{Diehl:1998kh}
for the light-front valence wavefunction. Here we consider
\begin{equation}
\psi_3(x_i,{\bf k}_i) = A e^{b^2\left[M^2 - \sum_{i=1}^3
\left( \frac{m^2+{\bf k}_\perp^2}{x}
\right)_i \right]}. \label{lcwf}
\end{equation}
The parameters, which are listed in Table~\ref{gaussian} together with a model for the running
$\alpha_s$ with the UV scale $\Lambda$ and IR freezing mass  $M_0$,
\be
\alpha_s(p^2)= \frac{4\pi}{(9\log((p^2+M_0^2)/\Lambda^2))},
\ee
 provide a reasonable fit to the  proton form factor ~\cite{Lepage:1979za}, which is shown Fig.~\ref{formfactorplot}.
\begin{table}
\begin{tabular}{|cc|c|}
\hline
Description & Variable & Value \\ \hline
Running of $\alpha_s$ & $\Lambda$ & 205 $\mbox{ MeV}$ \\
Freezing of $\alpha_s$ & $M_0$ & 1.05 $\mbox{ GeV}$ \\
Gaussian fall-off & $b$ & 4.85 \\
Quark mass & $m$ & 270 $\mbox{ MeV}$
\\ \hline
\end{tabular}
\caption{\label{gaussian} A workable parameter set for the Gaussian
valence quark model }
\end{table}
These wavefunctions, however,  are too soft to be used in
conjunction with an impulse approximation for
 DVCS at sufficiently large $t$, were
  the $J=0$  can be extracted  experimentally.
This is illustrated in Fig.~\ref{fig:softformfactor} where it is seen
how the magnetic form factor in the impulse approximation with
these soft wavefunctions departs from its experimental value already
at momentum transfer of the order of  $3\mbox{ GeV}^2$.
Still it is illustrative as a benchmark, and probably not too misleading
in the interval  $1 < -t < 3 \mbox{ GeV}^2$, to plot the $1/x$ from factor,
$F_{1/x}(t)$. This is shown in Fig.~\ref{fig:1overxlightfront}.
Should one wish to extend the computation to somewhat higher $-t$
without abandoning the impulse approximation, a hard component with
a power-law falloff  would have to be  included in the wave
function.  Reasonable power law models are available in the
literature \cite{Schlumpf:1992pp,Brodsky:1994aw,deTeramond:2005su}.

It is worth making some generic observations about the
valence quark model. The first is that the Compton amplitude has no
imaginary part at LO and leading twist, because $H$ is
real, and in the model,  at the break-points $x=-\xi$ and
$x = \xi$,  $H(\xi,\xi,t)$
and $H(-\xi,\xi,t)$ vanish.   Finiteness of the GPD at the break points
 is a generally assumed feature.  This behavior is reminiscent of, for example the pion
distribution amplitude at the end-points,  and indeed arises
from the same underlying assumption, namely that the end-point region
(or break-point region in the case of GPD's) is governed by the
one-gluon-exchange evolution. That this may not necessarily be the
case was discussed for example in Ref.~\cite{Hoyer:2002qg} and its
consequences for DVCS in Ref.~\cite{Szczepaniak:2007af}

Secondly valence models do not include large-$x$ tails of the sea
quarks. The contribution of the sea quarks have been modeled for
example in Ref.~\cite{Gardestig:2003jw,Diehl:1998kh},
 and the models seems to  fare well when compared with data available at current kinematics. However
the lack of a usable dynamical calculation of the sea quark component of
the proton wavefunction makes the number of ad-hoc parameters increase
with each new Fock subspace added.

\begin{figure}
 \includegraphics[height=2.5in]{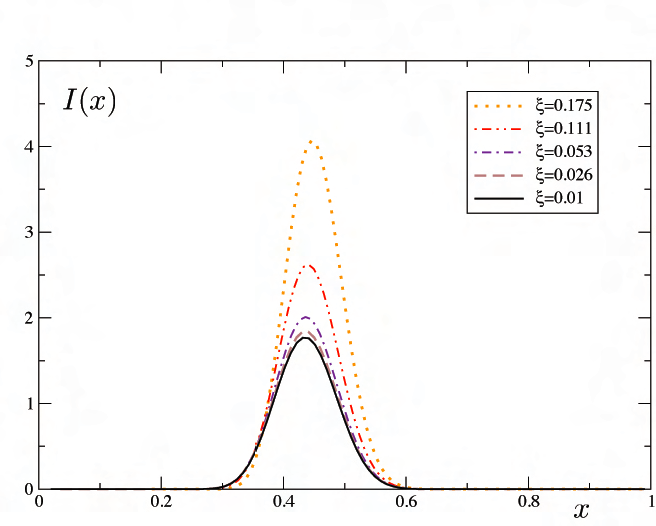}
\caption{\label{lightfrontfixed}
We plot the unintegrated handbag amplitude given  in Eq.(\ref{scaling}). The area under the left-most curve is smaller than all  others, and tags the $J=0$ fixed pole.}
\end{figure}

Still with our simplified valence version of the model we can see how the $J=0$
fixed pole arises here too. In Fig.~\ref{lightfrontfixed} we plot the unintegrated handbag amplitude
\begin{equation}
I(x) \equiv H(x,\xi,t)\left( \frac{1}{x-\xi} + \frac{1}{x+\xi}\right) \label{scaling}
\end{equation}
whose integral over $x$ yields the Compton amplitude.
(The relation between this $H$ GPD and that in the asymmetric frame as defined
in Refs.~\cite{Brodsky:2005vt,Vanderhaeghen:1999xj} is  $(1+\xi) H(x,\xi,t)=
H(x_{\rm asym},\zeta,t)$. )  As can be seen from
the figure, the area below the curve does not tend to zero in any limit.
The minimum of these areas is an irreducible contribution to the Compton
amplitude, independent of the skewness. It can be extracted by going to
the smallest possible skewness experimentally achievable.
We choose $t=-0.5\mbox{ GeV}^2$ for the plot, where
the fast exponential drop with $t$ has not yet set in, but this property
of there being a finite area independent of the skewness is not a
$t$-artifact.

\begin{figure}
 \includegraphics[height=2.5in]{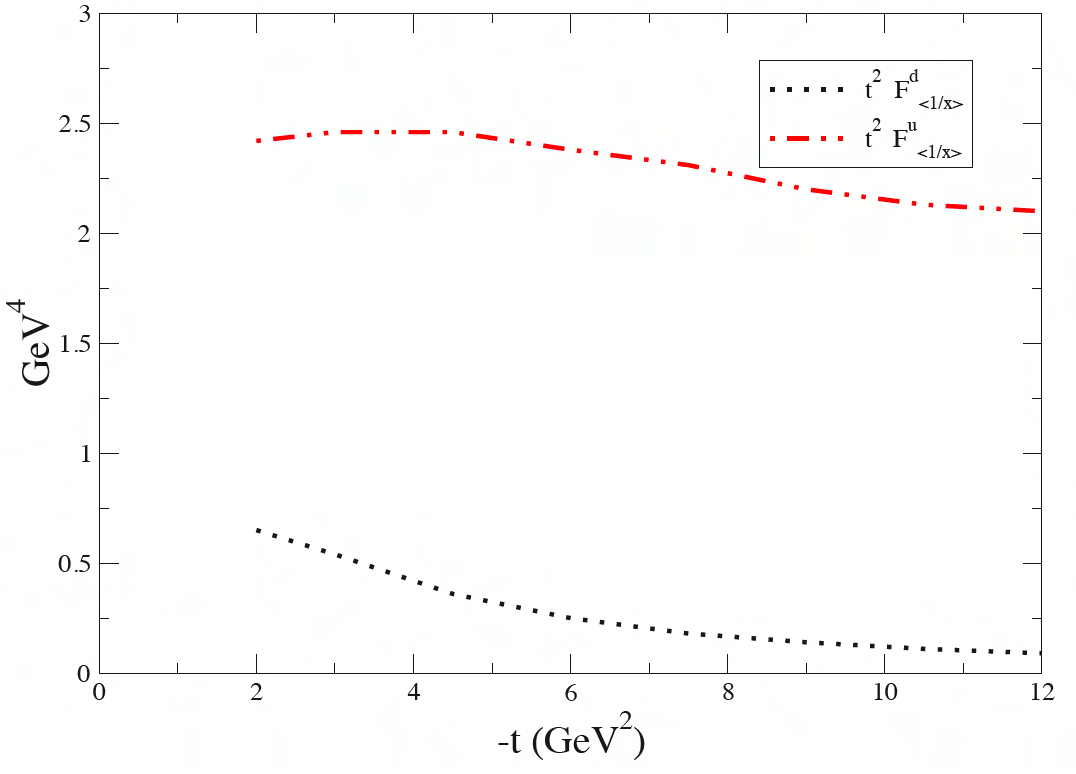}
\caption{\label{1overxDiehl}
An evaluation of the $1/x$ form factors of the proton has been carried
out in the Gaussian light-front constituent quark model, valid for large $-t$. The authors
present the flavor-separated form factors from a set of GPD's fit to a
number of conventional Dirac and Pauli form factors. Computer data from Ref.~\cite{Diehl:2004cx}.}
 \end{figure}

A more sophisticated fit to all available form factor data has been
carried on by Diehl et al.~\cite{Diehl:2004cx}. The fit includes Dirac
and Pauli form factors and a careful study of the propagation of their
errors. However the authors run into the Regge behavior of the pdf's at
small-x and therefore can only plot the $F_{1/x}$ at sizeable $t$. We
adapt their computation in Fig.~ \ref{1overxDiehl}.

\subsection{GPD in the covariant parton-nucleon model
\label{sec:parton-proton}}

The covariant parton model of Ref.~\cite{Landshoff:1970ff} was
extended to Compton scattering and DVCS in
Ref.~\cite{Brodsky:1973hm}. Since this model has not been widely
used in the past, we will rely on it more extensively for our
application, as a contribution to the current discussion on GPD's.
The idea is to construct a model of the quark-parton scattering
amplitude, with  unamputated quark legs of momentum $k$, $k'$ and
 Dirac indices $\alpha$,$\beta$ and two amputated proton legs with momentum
$p$, $p'$ and helicities $\lambda$, $\lambda'$.
Since the parton legs are off-shell, this amplitude is a function of
four different Lorentz scalars, that can be chosen as
the three Mandelstam invariants $\hat s=(p+k')^2$,
$t = \hat t=(k'-k)^2=(p'-p)^2$, $\hat u=(p-k)^2$
and $k^2$, The squared momentum of the returning parton can be expressed
as $k'^2=s+t+u-2M_N^2-k^2$. We will denote this amplitude by $T_{\lambda \lambda';
\alpha\beta}[\hat s,t,\hat u,k^2]$.
Once the dependence on nucleon helicity has been factored out the
 reduced amplitude can be expanded in the basis of Dirac matrices in the parton indices
 multiplied by scalar functions of the parton-nucleon Mandelstam variables.
 Eventually the $H$ generalized parton distribution can be expressed in terms of such scalar functions.
 \begin{equation} \label{GODpartonprotonrep}
H(x,\xi,t) = x p^+ \int
\frac{d^4k}{(2\pi)^4}
\delta\left(x p^+ -k^+\right) T[\hat s,t,\hat u,k^2]
\end{equation}
where  $p^+$  is the longitudinal momentum of the target.
For example, a model in which the parton nucleon amplitude
 is  taken to be described be an $\hat u$ -channel exchange
of a diquark of mass $\lambda$ would correspond to
\begin{equation} \label{simplediquark}
T[\hat s,t,\hat u,k^2]=(ig(k^2))
\frac{1}{(p-k)^2-\lambda^2+i\epsilon}
(ig(k^{'2}))
\end{equation}
with vertex functions $g(k^2)$ describing off-shell partons. These
  are expected to become perturbative at  large parton virtuality $k^2$ and at small virtuality are expected to be  soft {\it e.g.} determined by
the constituent quark mass, $m_q$.  These features can be incorporated by
writing a dispersion representation for $g(k^2)$  in the form
\begin{equation} \label{fullvertex}
g(k^2) =  \int_0^\infty d\Lambda^2
\frac{\rho(\Lambda^2)}{k^2-\Lambda^2+i\epsilon}
\end{equation}
with the spectral function given by
\begin{equation}
\rho(\Lambda^2)=  g m_q^4 \frac{d}{d\Lambda^2}\delta\left(  \Lambda^2-m_q^2
\right)\ .
\end{equation}
The support of the resulting GPD's, as shown  below, is the standard
$- 1 < x <1$ region which arises in a perturbative analysis where
the parton-nucleon vertex function has no structure. In the DIS
limit, where one can think of partons as essentially free, this is
probably a good approximation, although it has been speculated in
Ref.~\cite{VanDyck:2007jt} which at the limited $Q^2$ accessible to
experiment, one might find
 GPD's with support outside of these nominal limits, which is due to
structure in the $k^-$ plane stemming from the vertex functions.
Given that solid theory would rely on controlling the effect of quark
confinement on the analytical structure of vertex functions, we
ignore this possibility here but leave the question open.
Consider therefore for now the model defined by Eqs.(\ref{simplediquark}),(\ref{fullvertex}).
We employ the variable $\tilde{k}^- = P^+k^-$ so
that an inverse power of $P^+$ comes out of the $k^-$ integral and yields
\begin{eqnarray}
\label{GPDpartonprotonrep}
H(x,\xi,t) & = &  (x+\xi)\int_0^\infty \frac{
d\arrowvert {\bf k}_\perp \arrowvert^2}{2(2\pi)^4} \int_0^{2\pi}d\phi_\perp \nonumber \\
& \times & \int_0^\infty d\tilde{k}^-  T[\hat s,t,\hat u,k^2]
\end{eqnarray}
in place of Eq.(\ref{GODpartonprotonrep})  and with $k^+$, $k'$
fixed as above. Finally, we introduce both $\hat s$ and $\hat u$
channel amplitudes with a relative factor $\gamma$. That is, we generalize
  Eq.(\ref{simplediquark}) to,
\begin{eqnarray}  \label{doublediquark}
 \frac{T(\hat s,t,\hat u,k^2,k^{'2})}{ (ig(k^2)) (ig(k^{'2}))}  & = &
  \frac{1}{(p-k)^2-\lambda^2+i\epsilon}  \nonumber \\
   & + &
\frac{ \gamma }{(p'+k)^2-\lambda^2+i\epsilon}
\end{eqnarray}
The crossed, $\hat s$-channel term is necessary because of the known
forward relation yielding conventional antiquark pdf's
$H(x<0,\xi=0,t=0)=-\bar{q}(-x)$, The  $\hat u$-channel amplitude
 yields a non-vanishing $H(x)$ for $x>\xi$ only \cite{Brodsky:2005vt}. Since the
valence part of the antiquark distribution functions are known to be much
smaller than the quark pdf's, the   factor $\gamma$ accounts for suppression of
 the $\hat s$-channel.
We recall that the $1/x$ from factor, which we are interested in is defined through the valence region.
If we were to write
a general representation for $T$ containing the sea contributions in the low-$x$ region
the $\hat s$ and $\hat u$ channel terms would become constrained by crossing symmetry.

\begin{figure}
 \includegraphics[height=2.2in]{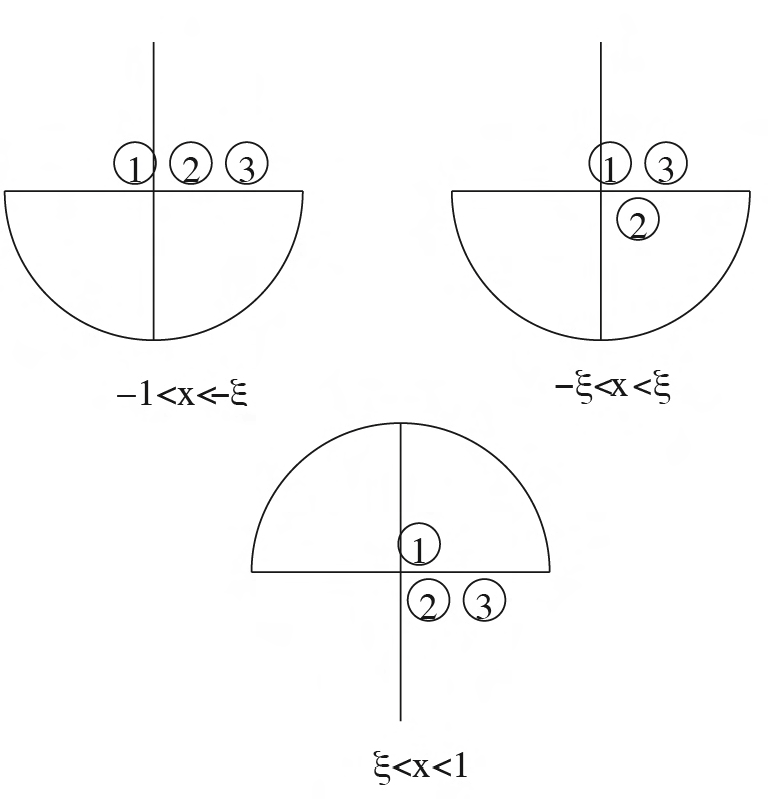}
\caption{\label{poleplacement}
Integration over the $\tilde{k}^-$  variable in Eq.(\ref{GPDpartonprotonrep}) is performed in the complex plane where
 location of the poles  of the integrand is determined by the denominators in Eq.(\ref{doublediquark}).
 Location of these poles is in agreement with Ref.~\cite{Diehl:1998sm}.
  For $x<\xi$,   $H(x<-\xi)$ receives a contribution
from the $\hat s$-channel and  for $x<\xi$, $H(x>\xi)$ is determined by the $\hat u$-channel.
}  \end{figure}

The parton-proton amplitude so defined is an holomorphic function of
$\tilde{k}^-$ and has four poles, three in each of the $s$ and $u$ channels, whose
positions are depicted in Fig.~\ref{poleplacement}.
The $\hat u$-channel diquark propagator yields a simple pole denoted as $\kappa^-_1$  and
given by the condition $\hat u-\lambda^2+i\epsilon=0$. The vertex functions
yield two double poles respectively, denoted by $\kappa^-_2$ for $k^2-\Lambda^2
+i\epsilon=0$ and $\kappa^-_3$ for $k^{'2}-\Lambda^2+i\epsilon=0$.
Finally the $s$-channel pole from $s-\lambda^2+i\epsilon=0$
is denoted $\kappa^-_4$. After deforming the $\tilde k^-$ integral, picking up the poles and
performing the differentiation with respect to $\Lambda$ ({\it c.f.} Eq.(\ref{fullvertex}))
on both propagators independently we  obtain~\cite{error}
 \begin{widetext}
\begin{eqnarray} \label{Hdiquark1}
H(-\xi< x < \xi,\xi,t)   &= &  - g^2m_q^8(x+\xi)
\int_0^\infty \frac{ d\arrowvert {\bf
k}_\perp\arrowvert^2}{2(2\pi)^3} \int_0^{2\pi} d\phi_\perp
\left[  \frac{1}{1+x} \left(
\frac{\gamma}{(k^2-m_q^2)^2(k^{'2}-m_q^2)^2}
\right)_{k^-=\kappa^-_{4}}  \right. \nonumber \\
&+ &
\left.  \left( \frac{1}{x+\xi}\right)^2 \frac{1}{((k^{'2}-m_q^2)^2\ar_{k^-=\kappa^-_2}}
\left[
\frac{x-1}{(u-\lambda^2)^2}  - \frac{\gamma(1+x)}{(s-\lambda^2)^2} +
\frac{2(x-\xi)}{k^{'2}-m_q^2} \left(
\frac{1}{u-\lambda^2}-\frac{\gamma}{s-\lambda^2}\right)
\right]_{k^-=\kappa^-_2} \right] \nonumber \\
\end{eqnarray}
and,
\ba \label{Hdiquark2}
H(\xi < x < 1,\xi,t)   &= &  g^2m_q^8 (x+\xi)
\int_0^\infty
\frac{ d\arrowvert {\bf k}_\perp\arrowvert^2}{2(2\pi)^3}
\int_0^{2\pi} d\phi_\perp
\left(\frac{1}{(k^{'2}-\Lambda^2)^2}\frac{1}{(k^2-\Lambda^2)^2}
\right)_{k^-=\kappa_1^-} \frac{1}{1-x} \nonumber \\
H( - 1 < x <-\xi),\xi,t)  & = & g^2m_q^8 (x+\xi)
\int_0^\infty
 \frac{ d\arrowvert {\bf k}_\perp\arrowvert^2}{2(2\pi)^3}
\int_0^{2\pi} d\phi_\perp
\left(\frac{1}{(k^{'2}-\Lambda^2)^2}\frac{1}{(k^2-\Lambda^2)^2}
\right)_{k^-=\kappa_4^-} \frac{1}{1+x}
\ea
\end{widetext}
which gives a positive (negative) definite $H$ function for $x>\xi$
(for $x<-\xi$) and the subscripts denote the value of $k^- =
\kappa^-_i$ obtained from the $i$-th pole. The perturbative scalar
diquark model presented here is, by construction ''hard''. That is,
it incorporates the asymptotic behavior based on dimensional
analysis, where $G_M(t)\propto 1/t^2$, and it is difficult to
reproduce the standard electromagnetic form factor at small momentum
transfer. However there exists a  parameter set which correctly
normalizes it to $F(0)=1$ and yields a Sachs form factor which is
never further from data than by a factor 1.5-2. The parameters are
$\Lambda=0.8\mbox{ GeV}$,   $m_q=0.4\mbox{ GeV}$, $g=25\mbox{ GeV}$
and $\gamma=0.2$.

\begin{figure}
 \includegraphics[height=2.5in]{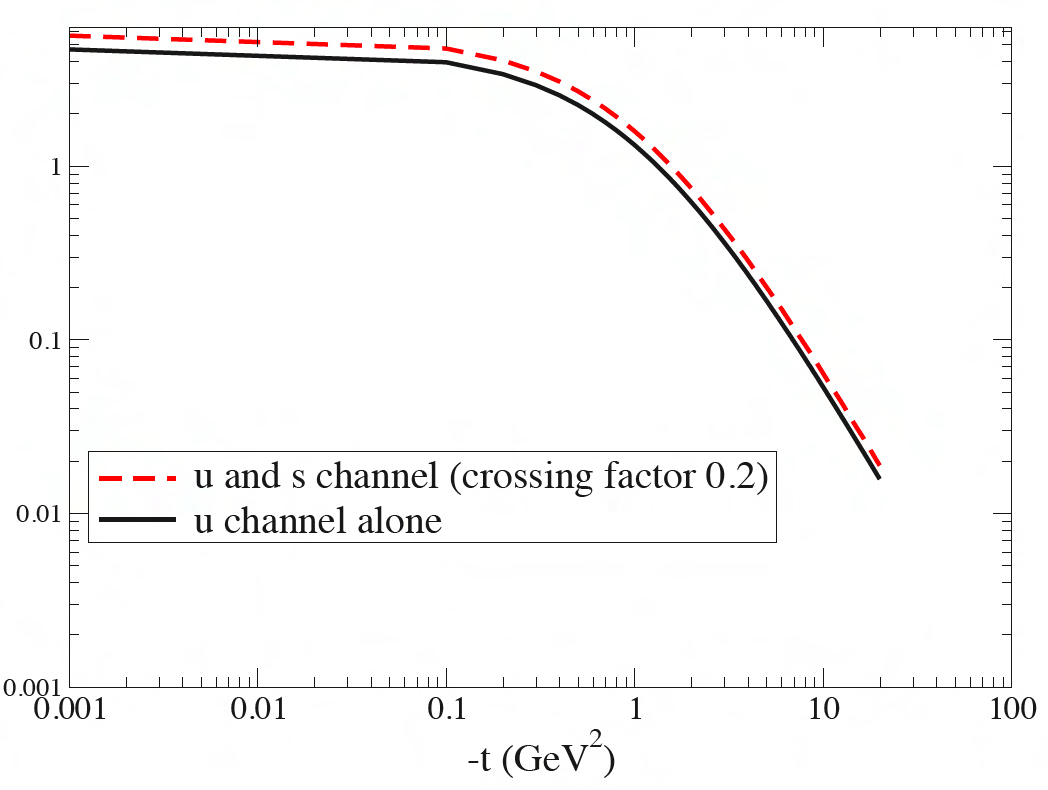}
\caption{\label{fig:oneoverxdiquark}
The $F_{1/x}$ form factor in the perturbative diquark model. At large $-t$ the form factor is power-law suppressed.}
\end{figure}

The resulting $1/x$ form factor is shown in Fig.~\ref{fig:oneoverxdiquark}.
To fix the isospin we have taken the rough
approximation $H^u=2H^d$ and weighted each with $e_q^2$ as would be
extracted from DVCS. More discussion on isospin can be found in the
appendix.
\begin{figure}
 \includegraphics[height=2.1in]{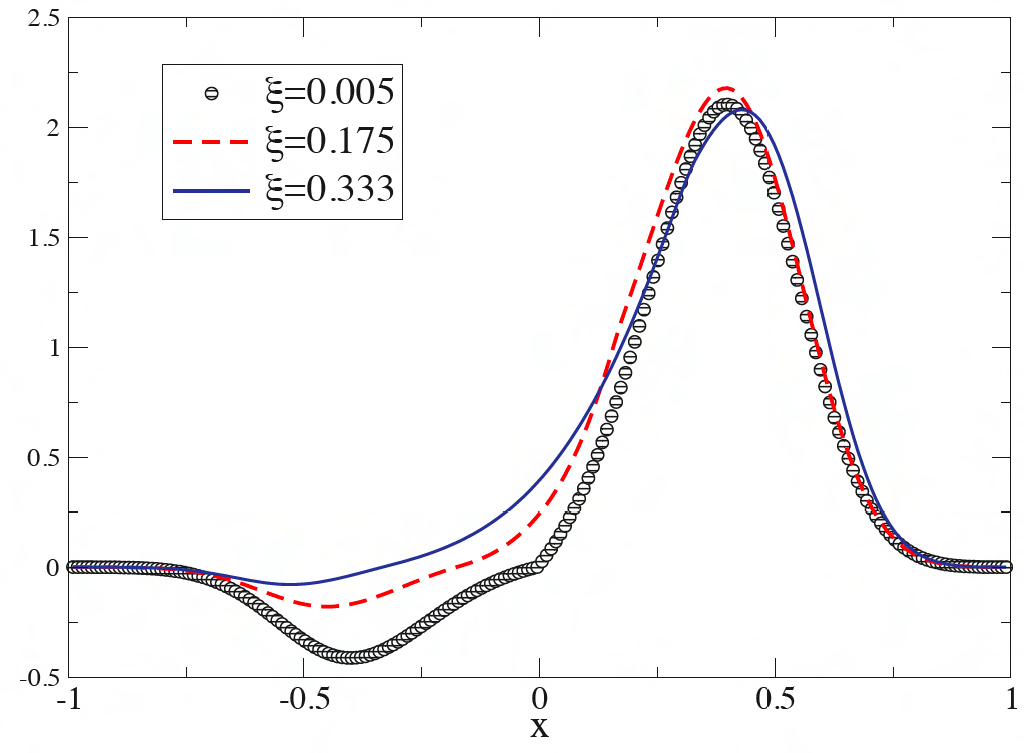}
\caption{\label{fig:gpddiquark}
 The $H(x,\xi,t)$  GPD  at fixed $-t=1 \mbox{ GeV}^2$ with the parameters discussed in text.
}
\end{figure}
In  Fig.~\ref{fig:gpddiquark} we plot $H(x)$ varying $\xi$. For
$x>0$, $H(x)$   describes  valence quarks and yields a finite $1/x$
form factor. For $x<0$ there is a small amount of antiquark sea (not
Regge-dominated) which also contributes. The figure  also
illustrates that  in this model $H(x)$
  vanishes at break-points, {\it i.e.} when $x\to \xi$.
The perturbative diquark model can be
 used as a template to provide a general description of the parton-proton scattering amplitude by means of Regge and spectral analysis, as advanced by
  in Ref.~\cite{Brodsky:1971zh}.

\section{Extracting the $J=0$ fixed pole from experimental data
\label{sec:expext}}

Early analysis of SLAC and Daresbury data revealed a fixed pole in the  real
proton and neutron Compton amplitudes.
The phenomenological analysis \cite{Damashek:1969xj,Dominguez:1972xe}
led to the following values for the forward limit of the $J=0$ pole amplitudes,
\begin{eqnarray}
 \label{exppole}
& & T_p^{J=0} = -3(1)  \frac{2M_N }{\alpha_{em.}} \mu b\mbox{ GeV} = -2.0(7), \nonumber \\
& &  T_n^{J=0} = -0.5(1)\frac{2M_N}{\alpha_{em.}}  \mu b \mbox{ GeV} = -0.3(7) \ .
\end{eqnarray}
for proton and neutron, respectively.
Earlier evaluation by Creutz {\it et al} \cite{Creutz:1968ds}
yielded  $T_p^{J=0} = -5(3)$ and it  was  observed  that in  magnitude and sign this amplitude
 is compatible with that of the Thomson term. It should be clear from
the dispersion analysis, however, that this is a coincidence,
and there is no reason why the amplitudes should be related.

A direct measurement of the fixed pole
has to date not been performed. We now sketch how it can be extracted from deeply virtual Compton scattering.

\subsection{Compton scattering}\label{expext:compton}
\begin{figure}
 \includegraphics[height=2.4in]{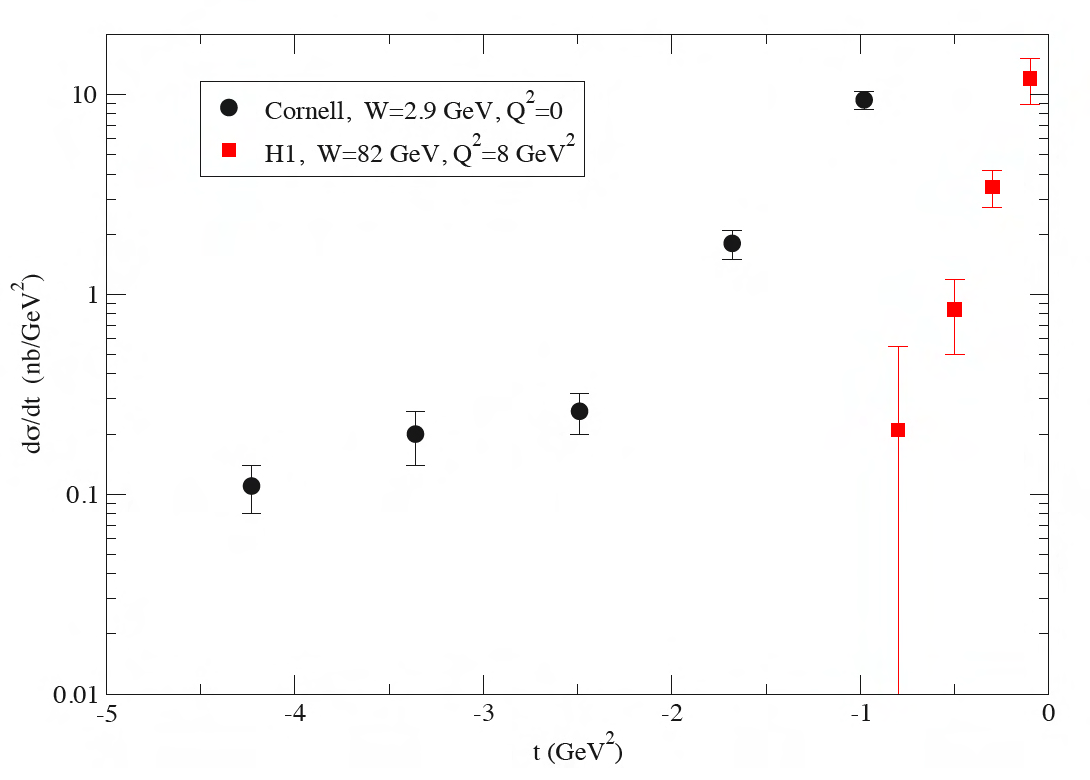}
\caption{\label{tdependenceCompton}
 The $t$ dependence of the real
Compton and deeply virtual Compton scattering cross sections from Cornell  \cite{Shupe:1979vg} and H1
\cite{Aktas:2005ty} experiments, respectively. }
\end{figure}

\begin{figure}
 \includegraphics[height=2.4in]{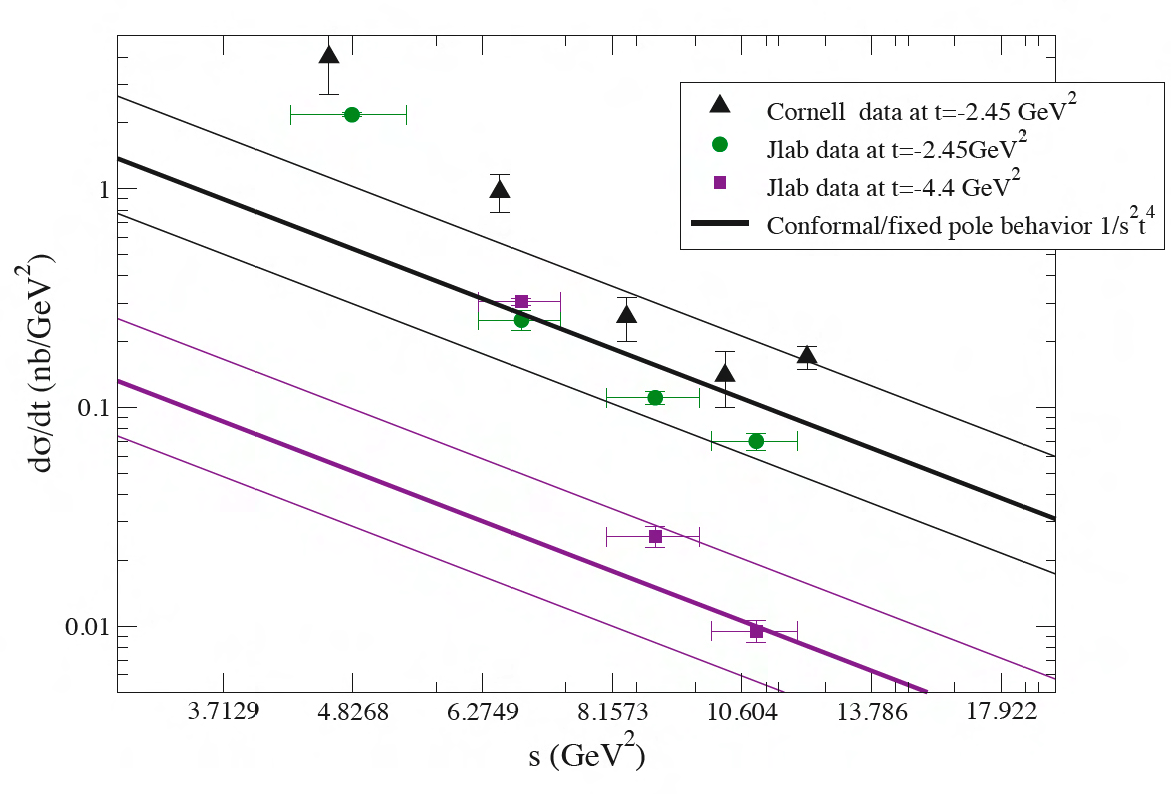}
\caption{\label{nofixedpoleyet}
The Cornell data from Ref.~\cite{Shupe:1979vg} is replotted for the fixed value of
$t=-2.45\mbox{ GeV}^2$. For comparison we also show recent JLab data from  Ref.~\cite{Danagoulian:2007gs}.}
\end{figure}

We shall comment here on existing data on Compton scattering in
regard to what they provide on  the extraction of the  $J=0$ pole.
First let us note that the DVCS data obtained at HERA is well in the
domain of Regge theory. At the stringently large $s$ and $Q^2$
required for the rest of its physics program, the experiments there
could not reach large $-t$ due to statistical limitations, and are
therefore dominated by leading Regge exchanges. This is apparent in
Fig.~\ref{tdependenceCompton} where we plot $t$-dependence of the H1
data. Fast exponential falloff is observed. For comparison
 we also show the data for real Compton. This high energy data has insufficient recoil, {\it i.e.} momentum transfer $t$
  and is dominated by the Pomeron. Note that QCD counting rules \cite{Brodsky:1974vy} predict a power-law falloff for
  exclusive process at fixed angle.
and at fixed angle, for real Compton scattering give,
\be
\frac{d\sigma}{dt} = \frac{f(\theta_{\rm CM})}{s^6}
\ee
However the law is
different and falls only as $s^{-2}$ in the Regge regime which has fixed-$t$ as opposed to fixed scattering angle.
This transition could be seen  in the Compton data from the Cornell
experiment and recent JLAB experiment, although typically pQCD
calculations are about an order of magnitude lower than the data
interpreted as fixed-angle Compton scattering \cite{Dixon}.
We shall now discuss $s$ dependence at fixed $t$. The prediction based on dominance of the
 $J=0$ pole is that the  differential cross section, $d\sigma/dt$  should falloff as $s^{-2}$.
 This is not yet clear from the Cornell data, which we plot
 in Fig.~\ref{nofixedpoleyet}  ignoring the highest
 $s$-point it appears, however,  the trend is correct and (in the log-log plot) the
slope seems to soften with increasing $s$, showing
the lessening influence of conventional Regge poles. Extending the
kinematic region in either $t$ or $s$ should help isolate the
asymptotic contribution expected to fall $s^{-2}$. The Jefferson Lab data
at the highest $t$ clearly fails to be consistent with the fixed pole form alone, but
the condition $s>>-t$ is not well satisfied. At slightly lower $-t$,
the data is consistent with the fixed-pole slope.
The lines in the figure correspond to a fit to the form
\be \label{FPfit}
\frac{d\sigma^{FP}_{\gamma p\to \gamma p}}{dt} = \frac{C^{FP}}{s^2 t^4}
\ee
where the central line and error bands correspond to a value $\log C^{FP} = 6.1(6)$
with $C^{FP}$ in units of  $nb\mbox{ GeV}^{10}$.

\begin{figure}
 \includegraphics[height=2.2in]{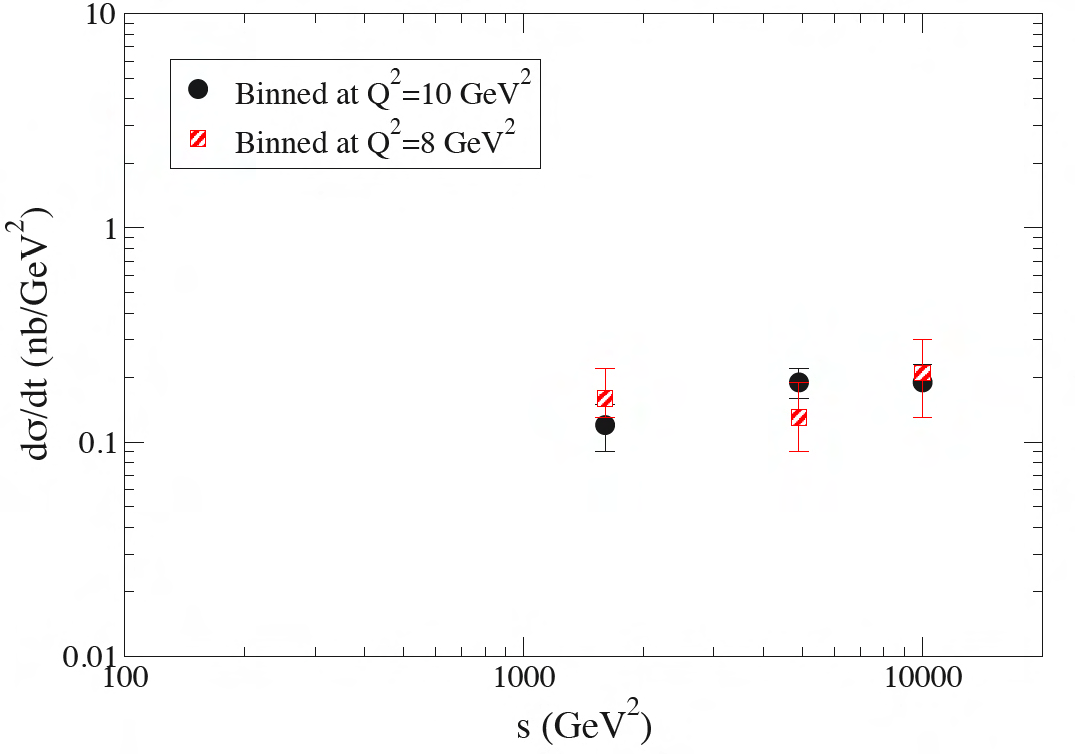}
\caption{
The H1 data from Ref.~\cite{newH1} for $t=-0.8GeV^2$.
 }\label{fig:newH1}
\end{figure}

The new H1 data at large energy \cite{newH1}, plotted in Fig. \ref{fig:newH1}
is dominated by the pomeron
even at their largest $-t=0.8\mbox{ GeV}^2$ bin (although the power-law exponent
has already diminished considerably from its hard-pomeron value), and
hence the fixed pole is not visible under this dominant Regge pole. Data
at larger $-t$ is needed.

If we cross the Compton amplitude $\gamma p\to \gamma p$ we obtain
the amplitude for proton-antiproton production by photofusion
$\gamma \gamma \to p\bar{p}$ that has been studied in  $e^-e^+$
collider experiments L3 at LEP \cite{L3ppbar} and Belle at
KEK~\cite{Belleppbar} among others.
  At fixed angle and large $s$, $t$, photon-photon annihilation into mesons  has been analyzed within
 perturbative QCD framework~\cite{brodskylepage1981,anselmino,anselmino2}.
The data, however, is again inconclusive in what pertains to the existence of
the $J=0$ fixed pole.

In Ref.~\cite{brodskylepage1981} it was shown that the photon-meson Compton scattering amplitude
 should have  the $J=0$ pole behavior.  In particular the Compton amplitude for vector mesons  is found to be
 \be
T_{\gamma V\to \gamma V}^{++}=
16\pi\alpha_{EM} F_{V}(t) (e_1^2+e_2^2)
\ee
However, the $J=0$ fixed pole decouples from pseudoscalar mesons, and
it is not likely that this relation will be tested soon.




\subsection{What to expect from future measurements}\label{subsec:future}
Given that the $J=0$ fixed pole cannot be claimed to have been
conclusively extracted  it would be very useful to have deeply
virtual Compton scattering data for, \be s>> Q^2 >>-t > -t_0 \sim
1\mbox{ GeV}^2. \ee Such kinematics is required for applicability of
Regge, and
 handbag-diagram phenomenology.

Currently, the HERA data does not satisfy the last inequality.
Jefferson Lab with a $12\mbox{ GeV}$ beam should be able to reach
 $s\simeq 40\mbox{ GeV}^2$, $Q^2 \simeq 6\mbox{ GeV}^2$, $t\simeq -3\mbox{ GeV}^2$
 and be able to measure the virtual Compton amplitude where the $J=0$ pole dominates
 and extract its form factor.
An electron-ion collider should be able to provide a definite
measurement in the challenging kinematics required to extract the
$J=0$ fixed pole and this adds to the further motivation of
considering such a machine.

\begin{figure}[h]
\includegraphics[height=1.2in]{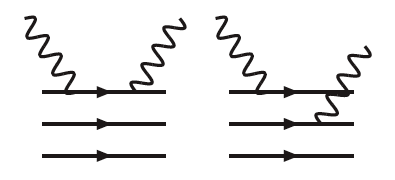}
\caption{In the handbag approximation incoherent scattering (left diagram) dominates  over coherent  processes
 with  all quarks participating.
}
\label{fig:whichquarkhit}
\end{figure}

An important test of the  handbag approximation and of whether the Compton
 amplitude  is dominated by the $1/x$ form factor, is to measure the
ratio of the differential cross-sections on the neutron and on the proton,
namely
\begin{equation}
R_{n/p} = \frac{\frac{d\sigma}{dt}(\gamma n\to \gamma
n)}{\frac{d\sigma}{dt}(\gamma p\to \gamma p)} \ .
\end{equation}
Assuming isospin symmetry, that is, $H^d_n=H^u_p$, the ratio becomes,
\begin{equation}\label{npratio}
R_{n/p} = \frac{\sum_n e_q^2}{\sum_p e_q^2} =
\frac{2e_d^2+e_u^2}{e_d^2+2e_u^2} = \frac{2}{3},
\end{equation}
if both photons couple to a single quark, as in the
handbag mechanism, (left diagram in Fig.~\ref{fig:whichquarkhit}), and
 smaller otherwise. In the extreme case
 of coherent scattering on valence quarks
(right diagram in Fig.~\ref{fig:whichquarkhit}), the ratio is expected to be close to,
\begin{equation}
R_{n/p} = \frac{\sum_n e_{q_i}e_{q_j}}{\sum_p e_{q_i}e_{q_j}} =\frac{1}{3}.
\end{equation}

Another interesting measurement would be to follow the $t$ dependence of the
Regge exponents, to ensure that indeed amplitudes at large
momentum transfer $t$ can be understood in as Regge exchanges with
intercepts $\alpha(t)\to -1$.  Jefferson lab could easily establish that
the Reggeons recede below $\alpha=0$ in
meson electroproduction, as the $t$ reach needed is only about $1\mbox{ GeV}^2$.
For each $t$ point, several $s$ measurements need to be taken with $s>>-t$
to check the $s^{-\alpha(t)}$ law.
As for $\rho^0_L$ electroproduction, there is abundant data on forward
production at small and moderate $Q^2$ and $s$ (see the compilation in Ref.~\cite{Bauer:1977iq}) and larger $Q^2$ \cite{Vanderhaeghen:1999xj}, but the
extraction of the $J=-1/2$ analogous to the Compton fixed pole requires, as commented above, higher
$t$.

\section{Conclusions \label{sec:last}}

The local coupling of two photons to the fundamental quark currents
of a hadron gives an energy-independent contribution to the Compton
amplitude proportional to the charge squared of the struck quark, a
contribution which has no analog in hadron scattering reactions. The
existence of this contribution, which is the analog of Thomson
scattering on the electrons of an atom at high energies, provides a
fundamental test of QCD. This paper is about the nucleon, but as we
have discussed in the introduction, the fixed-pole concept is
extensible to any bound system of pointlike (or effectively
pointlike) charges. We have shown that this local contribution has a
real phase and is universal, giving  the same contribution  for real
or virtual Compton scattering for any photon virtuality and skewness
at fixed momentum transfer squared $t$. The $t$-dependence of this
$J=0$ fixed Regge pole is parameterized by a yet unmeasured even
charge-conjugation form factor of the target nucleon.  The $t=0$
limit gives an important constraint on the dependence of the nucleon
mass on the quark mass through the Weisberger relation. Thus far,
contemporary fits using conventional parton distributions have
failed to unambiguously determine its value. Compton scattering of
real photons at large $t$ is especially interesting because since
the $J=0$ fixed pole gives a purely real, $s-$ and $Q^2-$
independent amplitude, the slowest-falling amplitude for any  hard
exclusive process in hadron physics.

The analysis of this paper provides a systematic procedure for
identifying and verifying the $J=0$ fixed pole contribution to real
and virtual Compton scattering. First, one identifies a candidate
$J=0$ contribution to the real Compton scattering cross section
$d\sigma/dt(\gamma p \to \gamma p) $ at $s
>> -t$ which scales as $1/s^2$ at fixed $t$; i.e., a contribution to
the Compton cross section which scales as the elementary
Klein-Nishina scattering cross section for $\gamma q \to \gamma q$
times the square of a form factor $F_{1/x}(t)$. Such a contribution
is possibly apparent in recent results from the E99-114 Hall A
experiment at J-Lab ~\cite{Danagoulian:2007gs}. If this contribution
is, in fact, due to the local coupling of the two photons to the
quark, it will be independent of the photon virtuality $q^2$ at
fixed $t$, when one measures high energy virtual Compton scattering
$\gamma^*(q) p \to \gamma p.$ Since the $J=0$ amplitude is real, it
will have maximum interference with the real Bethe-Heitler amplitude
in $e p \to e p \gamma$. This program should be practical at the
$12\mbox{ GeV}$ Jefferson Laboratory facility.

We have also discussed how the $ J=0$  fixed pole and the $1/x$ form
factor can be extracted from deeply virtual Compton scattering at
large $t$ and have examined predictions given by models of the $H$
generalized parton distribution.
The $J=0$ contribution are readily identifiable in DVCS at high
$-t<0.6-1\mbox{ GeV}^2$, where conventional Regge trajectories have
receded. One can then test specific models such as the diquark model
or  quark model with light front hadron wavefunctions, AdS/QCD
predictions, and lattice calculations.
%

We also note that the $J=0$ fixed pole appears as a local
energy-independent real contribution to the Compton amplitude for
other two-photon processes such as the timelike real and virtual
exclusive reactions $\gamma \gamma \to H \bar H$  $\gamma \gamma^*
\to H \bar H$ or $\bar{p}p\to H$ \cite{pinfold}.


\acknowledgments

We are indebted to many colleagues, among them  Mischa Gorchteyn,
Michael Peskin, Sidney Drell, Tim Londergan, Paul Hoyer, Anatoly
Radyushkin, Ivan Schmidt, Dieter Mueller and Markus Diehl, for
useful conversations. F.J. Llanes-Estrada warmly thanks the
hospitality of the
 SLAC National Accelerator Laboratory  theory group and the Indiana University Nuclear Theory Center where a
sizeable part of this work was completed, and partial
financial support from a Fundacion del Amo-Univ. Complutense fellowship.
as well as grants FPA 2004-02602, 2005-02327,
PR27/05-13955-BSCH (Spain)  and
 DE-FG0287ER40365 from the US Department of Energy (USA). Preprint SLAC-PUB-13478.

\appendix

\section{Origin of the $J=0$ fixed pole: a simple model\label{intropole}}

In Regge theory, a Regge pole at $J=\alpha(t)$ (a singularity of the
scattering matrix in the complex angular momentum plane) leads
to a high-energy behavior of the scattering amplitude proportional to $s^{\alpha(t)}$ for $s >> -t$.
A $J=0$  pole  thus corresponds to a scattering  amplitude which is energy-independent in the region of  momentum transfer for which $\alpha(t)= 0$.
As we have noted in the introduction, this contribution is a fundamental prediction  of QCD arising from the local two-photon interactions with the quark currents.

\begin{figure}
\includegraphics[scale=0.6]{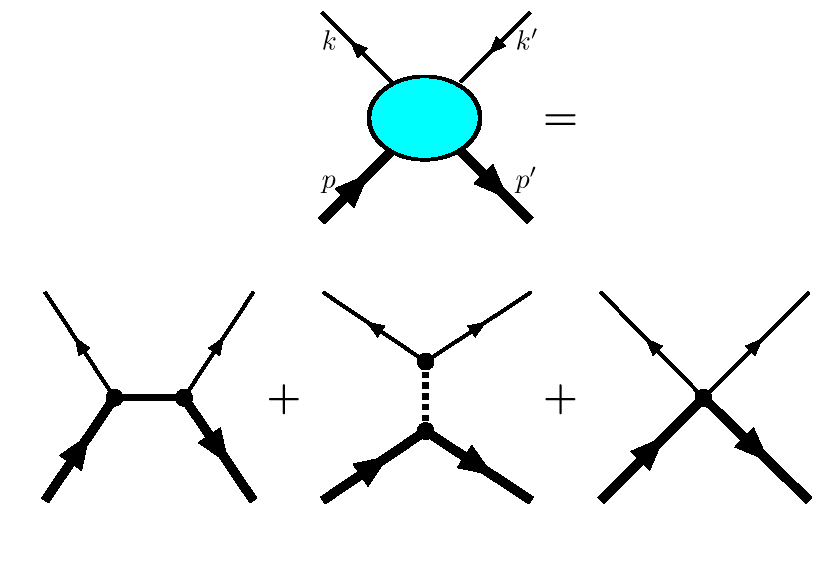}
\caption{\label{partonnucleon}
A simple, perturbative  model for a two-body scattering amplitude, given by the sum of an $s$-channel, $t$-channel and contact interaction.
%
}
\end{figure}

In perturbation
theory, energy-independence arises from contact interactions,
 as indicated by the last diagram in Fig.~\ref{partonnucleon}. To see this, consider an $s$- channel exchange
(lower-left panel on Fig.~\ref{partonnucleon}) of a spin-0 particle of mass $M$ (for simplicity we ignore the natural width). The corresponding amplitude $A_s$, is proportional to
 \begin{equation}
 A_s(s,M) = \frac{M^2}{M^2 - s }.  \label{s-channel}
 \end{equation}
The two limits are interesting. If $s\to \infty$ at fixed $M$, one has $A_s \sim s^{-1}$,
or $\alpha=-1$. We have normalized $A_s$  such that in the other limit,
$M \to \infty$, which corresponds to a point-like interaction with $\alpha=0$,
$A_s$ remains finite.  This $s$-channel exchange can be represented as
an infinite series of $t$-channel exchanges of different spins.
A standard way to expose this duality is to perform the Mellin transformation
which enables to write the amplitude in Eq.(\ref{s-channel}) as,
 \begin{equation}  \label{simpleamplitude}
 A_s(s.M) = \frac{1}{2\pi i} \int_{c -i\infty}^{c + i\infty}
 d\alpha \frac{\pi}{\sin\pi\alpha} \left(-\frac{M^2}{s}\right)^\alpha
 \end{equation}
where $0 < c < 1$.
For large c.m. energies $s > M^2$ the contour for the integral
can be closed to
encircle the positive real axis  with $\alpha > c$ and then
replaced by the sum over poles of $\sin\pi\alpha$ which occur at
integer  $\alpha = J$ with $J \ge 1$. The residues at these poles are
\begin{equation}
\mbox{Residue}_{\alpha = J}\left( \frac{\pi}{\sin\pi\alpha} \right)=(-1)^J
\end{equation}
and the amplitude becomes
\begin{equation}
A_s(s >  M^2 ) = \sum_{J \ge 1} (-1)^{J+1} \left(-\frac{s}{M^2}\right)^{-J}.
\end{equation}
For small c.m. energies, on the other hand,
$s<M^2$ the contour can be closed and the integral
replaced by a sum over poles to the left of  the $\alpha=c$ line the occur at
integer  $\alpha = -J$ including the  $J=0$ pole,
\begin{equation}
A_s(s <  M^2,M ) = \sum_{J \ge 0}(-1)^J\left(-\frac{s}{M^2}\right)^J
\label{low}
\end{equation}
(these relations are easy to check since they simply reconstruct
as a geometric series Eq.(\ref{s-channel}) )
The $s^J$ dependence of  individual amplitudes on the {\it r.h.s} of
Eq.(\ref{low})  is what is expected from exchange of a spin-$J$ object in the
$t$-channel. The large-$s$ behavior of the amplitude in Eq.(\ref{s-channel}) then
corresponds formally, in the $t$-channel, to a sum over exchanges of negative
spin.  Phenomenologically, hadron amplitudes with  $J=\alpha<0$ indeed occur
for physical $s$ and large and negative momentum transfer
$t < t_0 < 0$ \cite{Coon:1974wh}.
This is analogous to the simple model defined by the amplitude in
Eq.(\ref{s-channel}), in the language of Regge phenomenology, where
the  asymptotic  behavior  for $s >> M^2$  would correspond to  an
exchange of $J=\alpha=-1$ $t$-channel trajectory.  From Eq.(\ref{low})
it follows that the point-like interaction,
obtained in the limit $M^2 \to \infty$,
in the  Regge language, corresponds to an exchange of an object with spin, $J=0$.
We finally note that the presence of point-like scattering
is a necessary but not sufficient condition for the $J=0$ pole.
A combination  $A_s(s,M_1) - A_s(s,M_2)$ in the point like limit
$s<< M^2_1,M^2_2$ has a vanishing   $J=0$ amplitude~\cite{Creutz:1973zf}.
If either the parton-proton or parton-photon interactions have a pointlike
contact interaction, this will survive convolution with the rest
of the amplitude and reflect as a $J=0$ component of the photon-proton
amplitude. Conversely, experimentally establishing this
becomes a signature of point-like scattering on underlying
  elementary  constituents.

\section{Isospin analysis\label{sec:isospin}}

Throughout the paper quark charges are measured in units of the electron charge. The quark flavor decomposition of the proton GPD is given by
\begin{equation}
H^p_{F_1}= \sum e_q H^{q/p} = \frac{2}{3} H^{u/p} - \frac{1}{3}H^{d/p}.
\end{equation}
We ignore the quark sea, and assume the naive quark
model assignment $p=uud$, $n=udd$, and therefore set
$H^{u/p}=2H^{d/p}$ and neglect $H^{s/p}$
We will call $H^{d/p}$ simply $H$ and therefore
\begin{equation}
H^p_{F_1}= H. \\ \\ 
\end{equation}
In the case of the neutron, and profiting from isospin symmetry, we have
\begin{equation}
H^n_{F_1}= \frac{2}{3} H^{u/n} - \frac{1}{3}H^{d/n}=
\frac{2}{3} H^{d/p} - \frac{1}{3}H^{u/p}=0,
\end{equation}
which of course is expected to receive corrections
from sea quarks.
Turning to DVCS, the relevant combinations are now
\begin{eqnarray} \label{DVCSflavor}
H^p_{DVCS} &= &  \sum e_q^2 H^{q/p} =\frac{4}{9} H^{u/p} + \frac{1}{9}H^{d/p}=H
 \nonumber \\
H^n_{DVCS} &= &  \frac{2}{3}.
\end{eqnarray}
A simultaneous analysis of DVCS for the proton and the neutron
allows the extraction of both $u$ and $d$ $1/x$ moments \ba \int
\frac{dx}{x} H^u(x,0,t) = \frac{3}{5}\left( 4
F^p_{1/x}(t)-F^n_{1/x}(t) \right) \\ \nonumber \int \frac{dx}{x}
H^d(x,0,t) = \frac{3}{5}\left( - F^p_{1/x}(t)+4F^n_{1/x}(t) \right),
\ea which extrapolated to $t\to 0$ can be compared with
Eq.(\ref{Weisberger}).



\end{document}